\documentstyle[12pt,epsfig]{article}
\begin{document}

\title{Magnetoconductance Autocorrelation Function \\
for Few--Channel Chaotic Microstructures}
\author{\\ \\P.-B. Gossiaux\\
National Superconducting Cyclotron Laboratory,\\
Michigan State University, East Lansing, \\
Michigan 48824, U.S.A.\\ \\
Z. Pluha\v r\\
Charles University, Prague, Czech Republic\\ \\
and\\ \\
H.A. Weidenm\"uller\\
Max-Planck-Institut f\"ur Kernphysik, \\
D-69029 Heidelberg, Germany\\ \\ \\}
\date{\today}

\maketitle


\begin{abstract}

Using the Landauer formula and a random matrix model, we investigate the
autocorrelation function of the conductance versus magnetic field strength
for ballistic electron transport through few-channel microstructures with
the shape of a classically chaotic billiard coupled to ideal leads. This
function depends on the total number $M$ of channels and the parameter $t$
which measures the difference in magnetic field strengths. Using the
supersymmetry technique, we calculate for any value of $M$ the leading terms
of the asymptotic expansion for small $t$. We pay particular attention to
the evaluation of the boundary terms. For small values of $M$, we
supplement this analytical study by a numerical simulation. We compare our
results with the squared Lorentzian suggested by semiclassical theory and
valid for large $M$. For small $M$, we present evidence for non--analytic
behavior of the autocorrelation function at $t = 0$.


\end{abstract}

\newpage

Recent advances in semiconductor technology have made it possible to study
experimentally the transport of ballistic electrons through microstructures
\cite{ree89,bee91} with the shape of a classically chaotic billiard. Two
aspects of the dependence of the conductance $G(B)=(e^{2}/h)g(B)$ on an
external magnetic field $B$ have received \cite{Mar92,cha94} particular
attention: The suppression of weak localization at small values of $B$,
and the form of the conductance autocorrelation function $C(\Delta B) =
\overline{\delta g(B)\delta g(B+\Delta B)}$. Here, $\delta g(B) = g(B) -
\overline{g(B)}$ is the difference between the dimensionless conductance
$g(B)$ and its mean value. The bars indicate an average which experimentally
is taken over the Fermi energy or the applied gate voltage.
 
In a previous paper \cite{Plu95}, we have calculated the generic form of the
weak localization peak. In the present paper, we address the generic form of
the conductance autocorrelation function $C$. For ideal coupling between
leads and microstructure, this function depends on the number $M$ of channels
in the leads, and on $t$, a measure of $\Delta B$ to be defined below.

Several authors have calculated the autocorrelation function $C$ in an
approximate way. In the semiclassical approximation \cite{jal90}, $C$ was
found to have the form of a squared Lorentzian. The semiclassical limit is
expected to apply only when $M \gg 1$. Moreover, the non--diagonal terms
in the Gutzwiller trace formula are neglected. Attempts to determine $C$
from a random matrix model for the scattering matrix with the help of a
Brownian motion model \cite{rau95,frp95} have not been successful. Use of
a random matrix model and the supersymmetry technique \cite{efe95,fra95}
has confirmed the squared Lorentzian shape of $C$. However, these were also
asymptotic calculations valid in the limit $M \gg 1$ and thus kin to the
semiclassical approximation. 

In the actual experiments \cite{Mar92,cha94}, the number of channels per
lead was quite small. Thus, it is not clear whether the theoretical results
just mentioned apply to this situation. Moreover, a theoretical study of the
full autocorrelation function is of considerable interest in its own right.
Indeed, $C$ is a four--point function, and to the best of our knowledge it
has not been possible so far to calculate this kind of function without
approximation analytically, using a random matrix model as starting point.
The work reported in this paper does not solve this problem fully either.
However, we succeed in calculating analytically the leading terms in an
asymptotic expansion of $C$ for small $t$. We believe that in so doing, we
display interesting and novel aspects of the supersymmetry technique,
especially in relation to the boundary terms. In addition, we are led
to question whether $C$ is analytic in $t$ near $t = 0$. 
 
Starting point is the random matrix model for the Hamiltonians of the chaotic
microstructure for two values of the external field. Our approach is very
close to that taken in Refs. \cite{efe95,fra95}, and also to our work on weak
localization \cite{Plu95}, except that we now have to deal with a four--point
function. We use the Landauer formula and Efetov's supersymmetry technique
\cite{Efe83} in the form of Ref. \cite{Ver85}. 

The calculation of the leading terms of the asymptotic expansion of
the autocorrelation function for small $t$ involves a generalization
of the usual polar coordinates used for two--point functions. The
transformation to these variables necessarily introduces the boundary
or Efetov--Wegner terms into the formalism. We devote particular
attention to these terms. To the best of our knowledge, integrals in
polar coordinates over graded vectors have first been treated by
Parisi and Sourlas \cite{par79}, and over graded matrices by Efetov
\cite{Efe83} and Wegner, see Ref. \cite{Ver85}. Our central result
given in Eq.~(\ref{int-split-4b}) below has first been derived in a
different way by Efetov \cite{Efe83} and by Zirnbauer \cite{zir86}. 
The derivation which we employ makes use of the method of boundary
functions suggested by Berezin \cite{ber87}. For a more general
discussion of coordinate transformations in graded integrals we refer
to the papers by Rothstein \cite{rot87}, and by Zirnbauer and Haldane
\cite{zir95}.

As a complement to the analytical work which yields only the leading terms
of the asymptotic expansion, we simulate the correlation function numerically.
We focus attention on small values of the channel number $M$.

In Section \ref{formulation}, we formulate our approach. The integral theorem
used in calculating the asymptotic expansion, is derived in Section
\ref{par-of-man}. The analytical calculation is discussed in Section
\ref{asy-exp}, our numerical results are presented in Section \ref{num-sim}.
Section \ref{conclusions} contains the conclusions. Additional mathematical
details are given in two Appendices.

\section{Formulation of the Problem}
\label{formulation}

Much of the development in this section is similar to that of ref. 
\cite{Plu95} where further details may be found. We try to keep this 
presentation both brief and self--contained.  

\subsection{Physical Assumptions}
\label{physicalassumptions}

We thus
consider ballistic electron transport through a two--dimensional 
microstructure with the shape of a classically chaotic billiard in the 
presence of an external magnetic field $B$. We wish to determine the 
generic features of the autocorrelation function 
\begin{equation}
\label{autocorrelation}
\overline{\delta g(B^{(1)}) \delta g(B^{(2)})} = \overline{g(B^{(1)}) 
g(B^{(2)})} - \overline{g(B^{(1)})} \ \overline{g(B^{(2)})}. 
\end{equation}
Here, $g(B)$ is the dimensionless conductance, and the bar denotes an 
average. Experimentally, this average is typically taken over the applied
gate voltage \cite{Mar92}, while theoretically we take it over a suitably
chosen random matrix model. We assume that the Hamiltonians $H^{(r)}$ of
the closed billiard exposed to the two fields $B^{(r)}$, $r = 1,2$, are
described by two ensembles of Hermitean random matrices $H^{(r)}_{\mu \nu}$
of dimension $N$ which have the form 
\begin{equation}
\label{hamiltonian}
H^{(1)}_{\mu \nu} = H^{(I)}_{\mu \nu} - \sqrt{\frac{t}{N}} H^{(II)}_{\mu \nu},
\\ \ \
H^{(2)}_{\mu \nu} = H^{(I)}_{\mu \nu} + \sqrt{\frac{t}{N}} H^{(II)}_{\mu \nu}.
\end{equation}
The matrices $H^{(C)}_{\mu \nu}$ with $C = I, II$ belong to two independent 
(uncorrelated) Gaussian unitary ensembles (GUE's) with first and second 
moments given by
\begin{equation}
\label{moments}
\overline{H^{(C)}_{\mu \nu}} = 0, \ \ 
\overline{H^{(C)}_{\mu \nu} H^{(C)}_{\mu' \nu'}} 
= \frac{\lambda^2}{N} \delta_{\mu \nu'} \delta_{\mu' \nu}.
\end{equation}
The parameter $\lambda$ fixes the mean level spacing $d$ of each ensemble 
and must be adjusted to the local mean level spacing at the Fermi energy 
$E_F$ of the billiard. In the limit $N \rightarrow \infty$ always 
considered in this paper, the average level density $d^{-1}$ takes the 
shape of Wigner's semicircle, and the Fermi energy $E_F = 0$ is chosen 
to lie at its center where $d = \pi \lambda / N$. As shown in ref. 
\cite{Plu95}, the parameter $t$ is related to the area $A$ of the   
billiard by $\sqrt{t} = k |B^{(1)} - B^{(2)}| A / (2 \phi_0)$ where $k$ 
is a numerical constant of order unity and where $\phi_0 = hc/e$ is the 
elementary flux quantum.

For this model to be valid, the fields $B^{(r)}$ must obey the following
two conditions. (i) $B^{(r)}$ must be strong enough so that for both
$r = 1,2$, the crossover transition from orthogonal to unitary symmetry 
is complete. In practice this requires fields stronger than a few millitesla,
cf. Ref. \cite{Plu95}. (ii) The fields must be weak enough for a GUE 
description to be valid. In other words, the classical cyclotron radius 
must be large compared to the diameter of the billiard.

The billiard is attached to two quasi one--dimensional leads. In each lead, 
there are $M/2$ transverse modes which define the incoming and outgoing 
channels. These channels are labelled $a = 1, \ldots, M/2$ in lead one and 
$b = M/2 + 1, \ldots, M$ in lead two. With $|\mu>$ denoting a complete 
orthonormal set of basis states in the billiard, and $c$ the channels in 
either lead, we denote the coupling matrix elements between billiard and 
leads by $W_{c \mu}(E)$. We assume that these matrix elements obey 
time--reversal symmetry, $W_{c \mu}(E) = W_{c \mu}^*(E)$ so that the effect
of the magnetic field is confined entirely to the interior of the billiard.
The characteristic dependence on energy $E$ of the elements $W_{c \mu}(E)$ 
is typically very slow on the scale of the mean level spacing $d$, and is 
therefore neglected.

We write $g(B^{(r)}) = g^{(r)}$ and use for $g^{(r)}$, $r = 1,2$, the Landauer
formula for zero temperature, 
\begin{equation}
\label{landauer}
g^{(r)} = \sum_{a = 1}^{M/2} \sum_{b = M/2 + 1}^{M} \left( |S_{ab}^{(r)}|^2 
+ |S_{ba}^{(r)}|^2 \right).
\end{equation}    
The elements of the scattering matrix $S_{cd}^{(r)}$ are given by
\begin{equation}
\label{scattering}
S_{cd}^{(r)} = \delta_{cd} - 2 i \pi \sum_{\mu \nu} W_{c \mu} 
[D^{(r)}]^{-1}_{\mu \nu} W_{d \nu}   
\end{equation}    
where $D^{(r)}$ is the inverse propagator
\begin{equation}
\label{propagator}
D^{(r)}_{\mu \nu} = E \delta_{\mu \nu} - H^{(r)}_{\mu \nu} 
+ i \pi \sum_c W_{c \mu} W_{c \nu}.   
\end{equation} 
We must put $E = E_F$ and recall that $E_F = 0$. For $a \neq b$, we have 
\begin{equation}
\label{trace}
|S^{(r)}_{ab}|^2 = 4 \ {\rm Tr} \left( \Omega^a [D^{(r)}]^{-1} \Omega^b 
[D^{(r)\dagger}]^{-1} \right),
\end{equation} 
where the trace runs over the level index $\mu$, and where
\begin{equation}
\label{omegac} 
\Omega^c_{\mu \nu} = \pi W_{c \mu} W_{c \nu}.
\end{equation}
For later use, we also define
\begin{equation}
\label{omega} 
\Omega_{\mu \nu} = \sum_c \Omega^c_{\mu \nu} = \sum_c \pi W_{c \mu} W_{c \nu}.
\end{equation}
 
Because of the unitary symmetry of our matrix ensemble, the coupling 
matrix elements $W_{c \mu}$ appear in all ensemble averages only in the 
form of the unitary invariants $X_{cd} = \pi \sum_{\mu} W_{c \mu} W_{d \mu}$.
As in ref. \cite{Plu95}, we assume that $X_{cd} = X_c \ \delta_{cd}$ is
diagonal in the channel indices. The parameters $X_c$ appear in the final
expressions only via the ``transmission coefficients'' or ``sticking
probabilities'' $T_c$ given by 
\begin{equation}
\label{transmission}
T_c = 4 \lambda X_c (\lambda + X_c)^{-2} \ .
\end{equation}
With $0 \leq T_c \leq 1$, the transmission coefficients measure the strength
of the coupling between billiard and leads. Motivated by the absence of
barriers between billiard and leads, we put $T_c = 1$ for all $c$ (maximal
coupling). Our model then depends only on $M$ and on $|B^{(1)} - B^{(2)}|$,
i.e., on two parameters which are fixed by experiment. All dependence on
$E_F$ and on $\lambda$ disappears (the parameter $\lambda$ appears only in
the transmission coefficients $T_c$).

The ensemble average of $g^{(r)}$ is obviously the same as for the case
of a single GUE, $\overline{g^{(r)}} = M/2$ for $r = 1,2$. The calculation
of the autocorrelation function, Eq.~(\ref{autocorrelation}), therefore
reduces to that of the ensemble average $\overline{g^{(1)} g^{(2)}}$.
To find $\overline{g^{(1)} g^{(2)}}$, we use both analytical and numerical
methods. Our analytical study of the autocorrelation function presented in
this and in the following three sections is based on Efetov's supersymmetry
method \cite{Efe83} in the version of ref. \cite{Ver85}. We refer to the
latter paper as VWZ. The numerical calculation of the autocorrelation
function is discussed in Section \ref{num-sim}.

\subsection{Supersymmetry Approach}
\label{supersymmetry}

We deal with a four--point function. This necessitates some changes in the
formulation of VWZ.

The generating function $Z(J)$ is given by the graded Gaussian integral
\begin{equation}
\label{generatingfunction}
Z(J) = \int {\cal D} [\Phi] \exp \left( i \sum_{\alpha r p} 
< \overline{\Phi}_{\alpha r p}, [(D + J) \Phi]_{\alpha r p} > \right)
\end{equation} 
where we use the notation
\begin{equation}
\label{notation}
< F, G > = \sum_{\mu} F(\mu) G(\mu).
\end{equation}
The $8 N$ dimensional graded vector $\Phi$ has components 
$\Phi_{\alpha r p}(\mu)$. The supersymmetry index $\alpha$ distinguishes
ordinary complex (commuting) integration variables ($\alpha = 0$) and
Grassmannn (anticommuting) variables ($\alpha = 1$). Later, the two values
0 and 1 of the supersymmetry index $\alpha$ will also be denoted by $b$ and
$f$, respectively. As in subsection \ref{physicalassumptions}, the
index $r = 1,2$ refers to the two fields. The index $p = 1,2$ refers to the 
retarded and the advanced propagator as usual. Finally, $\mu = 1, \ldots, N$.
The adjoint graded vector is defined by 
$\overline{\Phi}(\mu) = \Phi^{\dagger}(\mu) s$, with 
\begin{equation}
\label{signfunction}
s_{\alpha r p, \alpha' r' p'} = (-)^{(1 + \alpha)(1 + p)} \
\delta_{\alpha \alpha'} \delta_{r r'} \delta_{p p'} \ .
\end{equation}
The $8 N$ dimensional graded matrix $D$ is the graded inverse propagator,
\begin{equation}
\label{gradedpropagator}
D = E\;{\bf 1}_8 \times {\bf 1}_N - H + i L \times \Omega \ ,
\end{equation}
with
\begin{equation}
\label{gradedhamiltonian}
H = {\bf 1}_8 \times H^{(I)} - \tau_3 \times \sqrt{\frac{t}{N}} H^{(II)} \ .
\end{equation}
The $8 N$ dimensional graded matrices $L$ and $\tau_3$ are given by
\begin{equation}
\label{matrices}
L_{\alpha r p, \alpha' r' p'} = (-)^{(1 + p)} \
\delta_{\alpha \alpha'} \delta_{r r'} \delta_{p p'}, \
[\tau_3]_{\alpha r p, \alpha' r' p'} = (-)^{(1 + r)} \
\delta_{\alpha \alpha'} \delta_{r r'} \delta_{p p'} \ .
\end{equation} 
The $8 N$ dimensional graded matrix $J = J(\epsilon)$, a 
function of $M^2$ ordinary variables $\epsilon^{(r p)}_{ab}$, has the form  
\begin{equation}
\label{epsilon}
J(\epsilon) = \sum_{r p} I^{(r p)} \times J^{(r p)}(\epsilon) \ ,
\end{equation}
where
\begin{equation}
\label{auxiliary1}
I^{(r p)}_{\alpha' r' p', \alpha'' r'' p''} = (-)^{(1 + \alpha')} \
\delta_{r r'} \delta_{p p'} \delta_{\alpha' \alpha''} \delta_{r' r''}
\delta_{p' p''} \ ,
\end{equation}
\begin{equation}
\label{auxiliary2}
J^{(r p)}_{\mu' \mu''}(\epsilon) = \pi \sum_{ab} W_{a \mu'} W_{b \mu''}
\epsilon^{(r p)}_{ba} \ .
\end{equation}
The integration measure is ${\cal D} [\Phi] = \prod_{\alpha r p \mu} 
{\rm d} \Phi_{\alpha r p}(\mu) {\rm d} \Phi^{*}_{\alpha r p}(\mu)$.
Performing the Gaussian integration, we obtain 
\begin{equation}
\label{generatingfunction1}
Z(J) = {\rm Detg}^{-1}(D + J) 
\end{equation}
where the symbol ${\rm Detg}$ stands for the graded determinant. 
Eqs.~(\ref{gradedpropagator}) to (\ref{auxiliary2}) 
show that the matrix $D + J$ consists of four diagonal blocks corresponding 
to the two different magnetic fields and to retarded and advanced propagators, 
respectively. The function $Z(J)$ factorizes accordingly. Differentiating 
the resulting expression with respect to the source parameters $\epsilon$
and averaging over the ensemble yields
\begin{equation}
\label{average}
\overline{ |S^{(1)}_{ab}|^2 |S^{(2)}_{a'b'}|^2 } = 
\frac{\partial^4}{\partial \epsilon^{(11)}_{ab} \partial \epsilon^{(12)}_{ba} 
\partial \epsilon^{(21)}_{a'b'} \partial \epsilon^{(22)}_{b'a'} } 
\overline{ Z(J)}|_{\epsilon = 0} \ .  
\end{equation}

For later developments, we note that $Z(J)$ possesses an important 
symmetry. We consider transformations $T: \Phi(\mu) \rightarrow T \Phi(\mu)$ 
which leave the bilinear form $\overline{\Phi}(\mu) \Phi(\mu)$ invariant. Such 
transformations satisfy the condition
\begin{equation}
\label{tmatrix}
T^{-1} = s T^{\dagger} s \ .
\end{equation}
The set of $8$ dimensional matrices $T$ forms a graded unitary group
$G = \mbox{U}(2,2/4)$ with compact
and non--compact bosonic subgroups. For $t = \Omega = J = 0$, the generating
function $Z(J)$ is invariant under this group.
  
\subsection{Saddle--Point Approximation}
\label{saddlepoint}

We average $Z(J)$ over the ensemble, apply the Hubbard--Stratonovich
transformation as usual to reduce the number of independent integration
variables, and use the limit $N \rightarrow \infty$ to define a
saddle--point approximation.

In Eq.~(\ref{generatingfunction}), we have to average the term
\begin{equation}
\label{exponential}
\exp \left( - i \sum_{\alpha r p} < \overline{\Phi}_{\alpha r p},
(H \Phi)_{\alpha r p} > \right) \ .
\end{equation}
Using Eq.~(\ref{hamiltonian}), we find that the average is the product of
two terms,
\begin{equation}
\label{average1}
\overline{\exp \left( - i \sum_{\alpha r p} < \overline{\Phi}_{\alpha r p},
({\bf 1}_8 \times H^{(I)} \Phi)_{\alpha r p} > \right)} = \exp \left( -
\frac{\lambda^2}{2N} {\rm trg} S^2 \right) \ ,
\end{equation}
\begin{equation}
\label{average2}
\overline{\exp \left( i \sum_{\alpha r p} < \overline{\Phi}_{\alpha r p},
(\tau_3 \times \sqrt{\frac{t}{N}} H^{(II)} \Phi)_{\alpha r p} > \right)} =
 \exp \left( -
\frac{t}{N} \frac{\lambda^2}{2N} {\rm trg} [(S \tau_3)^2] \right) \ ,
\end{equation}
where 
\begin{equation}
\label{bilinearform}
S_{\alpha r p, \alpha' r' p'}
= < \Phi_{\alpha r p}, \overline{\Phi}_{\alpha' r' p'} > 
= \sum_{\mu} \Phi_{\alpha r p}(\mu) \overline{\Phi}_{\alpha' r' p'}(\mu) \ .
\end{equation}
Collecting results, we have with $E=E_F=0$,
\begin{equation}
\label{averagegenerating}
\overline{Z(J)} = \int {\cal D} [\Phi] \exp \left( i {\cal L}(S)
+ i \sum_{\alpha r p} < \overline{\Phi}_{\alpha r p}, [(i L \times \Omega + J)
\Phi]_{\alpha r p} > \right) \ ,
\end{equation}
where 
\begin{equation}
\label{lagrangean}
i {\cal L}(S) = - \frac{\lambda^2}{2N} ( {\rm trg} S^2 + \frac{t}{N} 
{\rm trg} [(S \tau_3)^2] ) \ .
\end{equation}
The following steps are quite standard and are only sketched here. After
the Hubbard--Stratonovich transformation and integration over the 
variables $\Phi$, $\overline{Z(J)}$ is written as an integral over the
eight--dimensional graded matrices $\sigma$, 
\begin{eqnarray}
\label{averagegenerating1}
&&\overline{Z(J)} = \int {\cal D} [\sigma] \exp \left( - \frac{N}{2}
( {\rm trg} \ \sigma^2 + \frac{t}{N} {\rm trg} [(\sigma \tau_3)^2] ) \right)
\times \nonumber \\
&& \exp \left( - {\rm Trg} \ \ln \ [- \lambda
( \sigma + \frac{t}{N} \tau_3 \sigma \tau_3) \times {\bf 1}_N
+ i L \times \Omega + J ] \right) \ .
\end{eqnarray}
The symbol trg (Trg) stands for the graded trace over matrices of dimension 
$8 \ (8 N)$, respectively. The saddle--point condition yields for the 
saddle--point manifold $\sigma_G$ the result
\begin{equation}
\label{manifold}
\sigma_G = - i Q,\;\;\;Q = T^{-1} L T \ .
\end{equation}
The matrices $T$ obey Eq.~(\ref{tmatrix}) and belong to the coset space
$G/P$ defined by $G/P = \mbox{U}(2,2/4) / \mbox{U}(2/2) \times \mbox{U}(2/2)$,
where $P = \mbox{U}(2/2) \times \mbox{U}(2/2)$ is the subgroup of matrices
in $G$ which commute with $L$. After integration over the massive modes and
taking $N \rightarrow \infty$, $\overline{Z(J)}$ takes the form 
\begin{equation}
\label{averagegenerating2}
\overline{Z(J)} = \int {\cal D}[Q] \exp
\left( i {\cal L}_{\mbox{eff}}(Q) + i
{\cal L}_{\mbox{source}}(Q,J) \right) \ ,
\end{equation}
where
\begin{equation}
\label{effectivelagrangean1}
i {\cal L}_{\mbox{eff}}(Q) = \frac{t}{2} \ {\rm trg} \ [(Q \tau_3)^2]
- {\rm Trg} \
\ln \ [ i \lambda (Q + \frac{t}{N} \tau_3 Q \tau_3) \times {\bf 1}_N
+ i L \times \Omega ]
\end{equation}
and
\begin{eqnarray}
\label{effectivelagrangean2}
&& i {\cal L}_{\mbox{source}}(Q,J) = \nonumber \\
&&\quad
- {\rm Trg} \ \ln
\left[ {\bf 1}_8 \times {\bf 1}_N + 
\left( i \lambda (Q + \frac{t}{N} \tau_3 Q \tau_3) \times {\bf 1}_N
+ i L \times \Omega \right)^{-1} J \right].
\end{eqnarray}
Summing over the level index $\mu$, introducing the transmission
coeffients $T_c$ of Eq.~(\ref{transmission}) and recalling that $T_c = 1$
for all $c$, we obtain
\begin{equation}
\label{result}
\overline{g^{(1)} g^{(2)}} = \int {\cal D}[Q]
\ \mbox{detg}(1+QL)^{-M}
\ R(Q)
\ \exp \left( - \frac{t}{2} \ {\rm trg} [(Q \tau_3)^2]
\right) \ ,
\end{equation}    
where the source term $R(Q)$ has the form
\begin{eqnarray}
\label{sourceterm}
&&R(Q)
= \frac{1}{2} M^2 \ {\rm trg} [G^{(11)} G^{(22)}] {\rm trg} [G^{(21)} G^{(12)}]
\nonumber \\
&&\qquad\;\;\;\,
+ \frac{1}{4} M^4 \ {\rm trg} [G^{(11)} G^{(12)}] {\rm trg} [G^{(21)} G^{(22)}]
\nonumber \\
&&\qquad\;\;\;\,
+\frac{1}{4} M^3 \ {\rm trg} [ G^{(11)} G^{(12)} G^{(21)} G^{(22)}
+ G^{(11)} G^{(22)} G^{(21)} G^{(12)} ] \ ,
\end{eqnarray}
with
\begin{equation}
\label{auxiliary3}
G^{(rp)} = G^{(rp)}(Q) = (1 + QL)^{-1} I^{(rp)} \ .
\end{equation}
The autocorrelation function is a function of two parameters, the channel
number $M$ and the parameter $t$ for the magnetic field,
\begin{equation}
C(t,M)
=\overline{\delta g^{(1)}\delta g^{(2)}}
=\overline{g^{(1)}g^{(2)}}-\overline{g^{(1)}}\,\overline{g^{(2)}}\;,
\label{C-definition}
\end{equation}
with $\overline{g^{(1)}g^{(2)}}$ given by the graded integral (\ref{result}).
The graded matrices $Q$ can be parametrized in terms of 32 real variables,
half of them commuting, the others, anticommuting. In calculations of the 
average two--point function, one typically deals with a total of 16 
integration variables. Except for this increase in the number of variables 
and for the form of the source terms (which are, of course, specific to 
our problem), the form of our result in Eqs.~(\ref{C-definition})
and (\ref{result}) is quite standard. In spite of this similarity,
the increase in the number of variables renders a full analytical evaluation
of the graded integral (\ref{result}) very difficult. As pointed out in
the Introduction, our analytical work is restricted to the region of small
$t$. We evaluate the leading terms of the asymptotic expansion of $C(t,M)$
in powers of $t$. Progress in this calculation depends crucially on the
proper choice of the 32 variables used in parametrizing the matrices $Q$.

\section{Parametrization of the Manifold}
\label{par-of-man}

The integral (\ref{result}) extends over the manifold of $Q=T^{-1}LT$, where
$T$ stands for the graded matrices of the coset space
$\mbox{U}(2,2/4)/\mbox{U}(2/2)\times\mbox{U}(2/2)$.
Exponentiating the coset generators shows that the matrices $T$ can be
parametrized by two four-dimensional graded matrices $t_{12}$ and $t_{21}$,
\begin{equation}
T=\left(
\begin{array}{ll}
\sqrt{1+t_{12}t_{21}}&t_{12}\\
t_{21}&\sqrt{1+t_{21}t_{12}}
\end{array}\right),
\label{T-mat}
\end{equation}
related to each other by
\begin{equation}
t_{21}=k t_{12}^{\dagger} \ ,
\label{t12-unit}
\end{equation}
with $k=\mbox{diag}(1,1,-1,-1)$. (In presenting the matrices explicitly,
we assume that the row labels of matrices of dimensions 8, 4, and 2 are
$(p,\alpha,r)$, $(\alpha,r)$, and by $r$,  respectively, and that the
indices follow in lexicographical order. The matrices of dimensions 8
and 4 are presented in block form.)
The 32 matrix elements of $t_{12}$ and $t_{21}$
represent the Cartesian coordinates of $T$. The range of the 
commuting coordinates is defined by the condition that the ordinary part
of the matrix $1+t_{12}t_{21}$ be positive
definite (non--compact parametrization in Boson--Boson blocks, compact
parametrization in Fermion--Fermion blocks).
Following the arguments of section 5.5 of VWZ, we find that the
integration measure ${\cal D}[Q]$ is identical with the invariant
integration measure for integration over the coset space. 
In Cartesian coordinates, this measure takes the form 
\begin{equation}
\mbox{d}\mu(T(t_{12},t_{21}))=\prod_{\alpha r,\alpha'r'}
\mbox{d}[t_{12}]_{\alpha r,\alpha'r'}
\mbox{d}[t_{12}]_{\alpha r,\alpha' r'}^{\star}\;,
\label{mea-car}
\end{equation}  
where for ordinary complex $z=x+iy$, our $\mbox{d}z\mbox{d}z^{\star} =
2 i\mbox{d}x\mbox{d}y$.
We write $\mbox{d}\mu(T)$ for $\mbox{d}\mu(T(t_{12},t_{21}))$.
Despite the simplicity of the measure, the Cartesian coordinates are
not well suited for evaluating our integrals \cite{Efe83}.
Therefore, we follow Efetov and change to polar coordinates.

\subsection{Polar Coordinates}
\label{pol-coo}

As in VWZ, we express the matrices $t_{12}t_{21}$ and $t_{21}t_{12}$ in
terms of their common eigenvalue matrix $\epsilon$ and of the eigenvector
matrices $u_{1}$ and $u_{2}$,
\begin{equation}
t_{12}t_{21}=u_{1}\epsilon u_{1}^{-1},\;\;\;
t_{21}t_{12}=u_{2}\epsilon u_{2}^{-1} \ .
\label{t12t21-diag}
\end{equation}
The unitarity relation (\ref{t12-unit}) requires that both diagonalizing
matrices $u_{1}$ and $u_{2}$ belong to graded unitary groups of type
$\mbox{U}(2/2)$,
\begin{equation}
u_{1}^{-1}=u_{1}^{\dagger},\;\;\;
u_{2}^{-1}=k u_{2}^{\dagger}k \ .
\label{uv-unit}
\end{equation}
To specify the matrices $\epsilon$, $u_{1}$ and $u_{2}$ unambiguously,
we order the eigenvalues $\epsilon_{\alpha 1}$ and $\epsilon_{\alpha 2}$ 
by the magnitudes of their ordinary parts,
$\mbox{ord}|\epsilon_{b1}|<\mbox{ord}|\epsilon_{b2}|$,
$\mbox{ord}|\epsilon_{f1}|<\mbox{ord}|\epsilon_{f2}|$. Moreover, we assume 
that the diagonalizing matrices $u_{1}$, $u_{2}$ are elements of the cosets
$\mbox{U}(2/2)/\mbox{U}(1)\!\times\!\mbox{U}(1)\!
\times\!\mbox{U}(1)\!\times\!\mbox{U}(1)$,
given by
\begin{equation}
u_{1}=u_{1x}u_{1\beta}u_{1\gamma},\;\;\;
u_{2}=u_{2x}u_{2\beta}u_{2\gamma} \ .
\label{u12-fac}
\end{equation}
Here
\begin{eqnarray} 
&&u_{1x}\!=\!\exp\left(\!
\begin{array}{llll}
0&\xi_{1b}&0&0\\
-\xi_{1b}^{\star}&0&0&0\\
0&0&0&\xi_{1f}\\
0&0&-\xi_{1f}^{\star}&0\end{array}
\!\right)\!,
u_{2x}\!=\!\exp\left(\!
\begin{array}{llll}
0&\xi_{2b}&0&0\\
-\xi_{2b}^{\star}&0&0&0\\
0&0&0&\xi_{2f}\\
0&0&-\xi_{2f}^{\star}&0\end{array}
\!\right)\!,
\nonumber\\ \nonumber\\
&&u_{1\beta}\!=\!\exp\left(\!
\begin{array}{llll}
0&0&0&\beta_{11}\\
0&0&\beta_{12}&0\\
0&\beta_{12}^{\star}&0&0\\
\beta_{11}^{\star}&0&0&0\end{array}
\!\right)\!,
u_{2\beta}\!=\!\exp\left(\!
\begin{array}{llll}
0&0&0&i\beta_{21}\\
0&0&i\beta_{22}&0\\
0&i\beta_{22}^{\star}&0&0\\
i\beta_{21}^{\star}&0&0&0\end{array}
\!\right)\!,
\nonumber\\ \nonumber\\
&&u_{1\gamma}\!=\!\exp\left(\!
\begin{array}{llll}
0&0&\gamma_{11}&0\\
0&0&0&\gamma_{12}\\
\gamma_{11}^{\star}&0&0&0\\
0&\gamma_{12}^{\star}&0&0\end{array}
\!\right),
u_{2\gamma}\!=\!\exp\left(\!
\begin{array}{llll}
0&0&i\gamma_{21}&0\\
0&0&0&i\gamma_{22}\\
i\gamma_{21}^{\star}&0&0&0\\
0&i\gamma_{22}^{\star}&0&0\end{array}
\!\right)
\nonumber \\
\label{u12xbg}
\end{eqnarray}
are exponentials of coset generators specified by 8 commuting parameters
$\xi_{p\alpha}$, $\xi_{p\alpha}^{\star}$, and by 16 anticommuting parameters
$\beta_{pq}$, $\beta_{pq}^{\star}$, $\gamma_{pr}$ and $\gamma_{pr}^{\star}$,
with $\alpha=b,f$, and $p,q,r=1,2$.
We parametrize the matrices $u_{px}$ by their off-diagonal matrix elements
$x_{p\alpha} = \mbox{e}^{i \arg \xi_{p\alpha}} \sin | \xi_{p\alpha} |$.
We note that the matrices
$u_{1}^{-1}t_{12}u_{2}$ and $u_{2}^{-1}t_{21}u_{1}$ commute with the
eigenvalue matrix $\epsilon$, denote these two diagonal matrices by $\lambda$
and $\bar{\lambda}$, respectively, and write the matrices $t_{12}$ and
$t_{21}$ as products,
\begin{equation} 
t_{12}=u_{1}\lambda u_{2}^{-1},\;\;\;
t_{21}=u_{2}\bar{\lambda}u_{1}^{-1} \ .
\label{t12-ulamu}
\end{equation}
According to Eqs.~(\ref{t12-unit}) and (\ref{t12t21-diag}), the matrices
$\lambda$ and $\bar{\lambda}$ are related to each other and to the eigenvalue 
matrix $\epsilon$ by
\begin{equation}
\bar{\lambda}=k\lambda^{\star},\;\;\;
\epsilon=\lambda\bar{\lambda}.
\label{lam-prop}
\end{equation}  
Substituting Eq.(\ref{t12-ulamu}) in Eq.(\ref{T-mat}) yields
\begin{eqnarray}
&&T=U\Lambda U^{-1},\quad U=U_{x}U_{\beta}U_{\gamma},
\nonumber\\ \nonumber \\
&&U=\left(\begin{array}{ll}
u_{1}&0\\
0&u_{2}
\end{array}\right),
\quad
\Lambda=\left(\begin{array}{ll}
\sqrt{1+\lambda\bar{\lambda}}&\lambda\\
\bar{\lambda}&\sqrt{1+\bar{\lambda}\lambda}
\end{array}\right) \ ,
\label{U-Lam}
\end{eqnarray}
with
\begin{equation}
U_{w}=\mbox{diag}(u_{1w},u_{2w}),\;\;\;w=x,\beta,\gamma,\quad
\lambda=\mbox{diag}(\lambda_{b1},\lambda_{b2},\lambda_{f1},\lambda_{f2})\;.
\label{Uw-lam1}
\end{equation}
These equations define the desired polar parametrization of $T$.
Out of 32 polar coordinates, 24 reside in the matrix $U$ (parameters
$x_{p\alpha},\beta_{pq},\gamma_{pr}$, with $\alpha=b,f$ and  $p,q,r=1,2$,
and their complex conjugates), 8 in the matrix $\Lambda$ (parameters
$\lambda_{\alpha r}$, with $\alpha=b,f$ and  $r=1,2$, and their complex
conjugates).

For the assumed ordering of the eigenvalues $\epsilon_{\alpha r}$, the
range of the absolute values of the ordinary parts of the commuting polar 
variables is specified by the inequalities
\begin{eqnarray}
&&0<\mbox{ord}|\lambda_{b1}|<\mbox{ord}|\lambda_{b2}|<\infty \ , \nonumber \\
&&0<\mbox{ord}|\lambda_{f1}|<\mbox{ord}|\lambda_{f2}|<1 \ , \nonumber \\
&&0<\mbox{ord}|x_{p\alpha}|<1 \ .
\label{int-reg}
\end{eqnarray}
The phase angles of the ordinary parts are allowed to take on all values
between $0$ and $2\pi$.
The integration measure $\mbox{d}\mu(U\Lambda U^{-1})$ in polar coordinates
is obtained from the Berezinian \cite{ber87}. We get
\begin{equation}
\mbox{d}\mu(U\Lambda U^{-1})=\mbox{d}\mu(U)\mbox{d}\mu(\Lambda) \ .
\label{mea-pol}
\end{equation}
Here $\mbox{d}\mu(U)$ denotes the measure for integration over the
matrices $U$, a product of measures for integration over the matrices
$U_{x\beta}=U_{x}U_{\beta}$ and $U_{\gamma}$,
\begin{equation}
\mbox{d}\mu(U)=\mbox{d}\mu(U_{x\beta})\mbox{d}\mu(U_{\gamma}) \ .
\label{mea-U}
\end{equation}
We have defined
\begin{equation}
\mbox{d}\mu(U_{x\beta})
=\prod_{p\alpha}\mbox{d}x_{p\alpha}\mbox{d}x_{p\alpha}^{\star}
\cdot\prod_{pq}\mbox{d}\beta_{pq}\mbox{d}\beta_{pq}^{\star}
\cdot\prod_{p}
(1-2\beta_{p1}\beta_{p1}^{\star}\beta_{p2}\beta_{p2}^{\star}) \ ,
\label{mea-Uxbet}
\end{equation}
\begin{equation}
\mbox{d}\mu(U_{\gamma})
=\prod_{pr}\mbox{d}\gamma_{pr}\mbox{d}\gamma_{pr}^{\star} \ .
\label{mea-Ugam}
\end{equation}
The quantity $\mbox{d}\mu(\Lambda)$ denotes the measure for integration
over the matrices $\Lambda$,
\begin{equation}
\mbox{d}\mu(\Lambda)
=\prod_{\alpha r}\mbox{d}\lambda_{\alpha r}
\mbox{d}\lambda_{\alpha r}^{\star}\cdot
\prod_{\alpha}(|\lambda_{\alpha 1}|^{2}-|\lambda_{\alpha 2}|^{2})^{2}\cdot
\prod_{rr'}(|\lambda_{br}|^{2}+|\lambda_{fr'}|^{2})^{-2}\;.
\label{mea-Lam}
\end{equation}
Eq.~(\ref{mea-Lam}) shows that the measure $\mbox{d}\mu(U\Lambda U^{-1})$
contains a nonintegrable singularity. It is located at
$\mbox{ord}(|\lambda_{b1}|^{2}+|\lambda_{f1}|^{2})=0$, i.e., at the surface
of the domain of integration. This singularity causes the occurrence of an
additional term, the Efetov--Wegner term. 

\subsection{The Efetov--Wegner Term}
\label{int-for}

For pedagogical reasons, we first exemplify the treatment of the singularity
for a simpler case. Indeed, the same problem arises in the integration over
the four--dimensional matrices $T_{a}$ belonging to the coset space 
$\mbox{U}(1,1/2)/\mbox{U}(1/1)\times\mbox{U}(1/1)$. We return to the
full problem below.

The properties of the four--dimensional matrices $T_{a}$ are the same as
those of their eight--dimensional counterparts $T$. The coset space consists
of the matrices $T_{a}$ satisfying $T_{a}^{-1}=s_{a}T_{a}^{\dagger}s_{a}$,
now with $s_{a}=\mbox{diag}(1,1,-1,1)$, modulo the subgroup of matrices which
commute with $L_{a}=\mbox{diag}(1,1,-1,-1)$. (We label the rows of graded
matrices of dimensions 4 and 2 by the indices  $(p,\alpha)$ and $\alpha$,
respectively.) The matrices have the form
\begin{equation}
T_{a}=\left(
\begin{array}{ll}
\sqrt{1+t_{12}^{a}t_{21}^{a}}&t_{12}^{a}\\
t_{21}^{a}&\sqrt{1+t_{21}^{a}t_{12}^{a}}
\end{array}\right)
\label{T-mat-4}
\end{equation}
where the two-dimensional blocks $t_{12}^{a}$ and $t_{21}^{a}$
are related to each other by
\begin{equation}
t_{21}^{a}=k_{a}[t_{12}^{a}]^{\dagger},
\label{t12-unit-4}
\end{equation}
with $k_{a}=\mbox{diag}(1,-1)$.
The matrix elements of these blocks contain the 8 Cartesian coordinates
of $T_{a}$. Again, the range of the commuting coordinates is defined by the
condition that the ordinary part of $1+t_{12}t_{21}$ be positive definite.
We consider the integral 
\begin{equation}
{\cal I}=\int\!\!\mbox{d}\mu(T_{a})F(T_{a})
\label{int-4}
\end{equation}
of a function $F(T_{a})$ over the coset space with the measure
\begin{equation}
\mbox{d}\mu(T_{a})=\prod_{\alpha\alpha'}
\mbox{d}[t_{12}^{a}]_{\alpha\alpha'}
\mbox{d}[t_{12}^{a}]_{\alpha\alpha'}^{\star} \ .
\label{mea-car-4}
\end{equation}
Polar coordinates are introduced \cite{Efe83} in the same way as described
above. This yields $T_{a}=U_{a}\Lambda_{a} U_{a}^{-1}$. The anticommuting 
coordinates reside in the eigenvector matrix $U_{a}$, and the commuting
ones, in the ``eigenvalue matrix'' $\Lambda_{a}$. We have
\begin{equation}
U_{a}=\left(\!
\begin{array}{ll}
u_{1a}&0\\
0&u_{2a}
\end{array}
\!\right),
\;\;\;
\Lambda_{a}=\left(\!
\begin{array}{ll}
\sqrt{1+\lambda_{a}\bar{\lambda}_{a}}&\lambda_{a}\\
\bar{\lambda}_{a}&\sqrt{1+\bar{\lambda}_{a}\lambda_{a}}
\end{array}
\!\right),
\label{U-Lam-4}
\end{equation}
where $u_{pa}$ and $\lambda_{a}$ denote the two--dimensional matrices
\begin{equation}
u_{1a}=\exp\left(\!
\begin{array}{ll}
0&\gamma_{1a}\\
\gamma_{1a}^{\star}&0
\end{array}
\!\right),\;\;\;
u_{2a}=\exp\left(\!
\begin{array}{ll}
0&i\gamma_{2a}\\
i\gamma_{2a}^{\star}&0
\end{array}
\!\right),\;\;\;
\lambda_{a}=
\left(\!
\begin{array}{ll}
\lambda_{ba}&0\\
0&\lambda_{fa}
\end{array}
\!\right),
\label{u12-lam-4}
\end{equation}
and where $\bar{\lambda}_{a}=k_{a}\lambda_{a}^{\star}$. Moreover,
\begin{equation}
t_{12}^{a}=u_{1a}\lambda_{a}u_{2a}^{-1},\;\;\;
t_{21}^{a}=u_{2a}\bar{\lambda}_{a}u_{1a}^{-1} \ .
\label{car-pol-4}
\end{equation}
For the range of the ordinary parts of the commuting polar variables, we
have $0<\mbox{ord}|\lambda_{ba}|<\infty$, $0<\mbox{ord}|\lambda_{fa}|<1$.
The integration measure in polar coordinates is given by
\begin{eqnarray}
&&\mbox{d}\mu(U_{a}\Lambda_{a}U_{a}^{-1})
=\mbox{d}\mu(U_{a})\mbox{d}\mu(\Lambda_{a}) \ ,
\nonumber \\ \nonumber \\
&&\mbox{d}\mu(U_{a})
=\prod_{p}\mbox{d}\gamma_{pa}\mbox{d}\gamma_{pa}^{\star} \ ,
\nonumber \\ \nonumber \\
&&\mbox{d}\mu(\Lambda_{a})
=\prod_{\alpha}\mbox{d}\lambda_{\alpha a}\mbox{d}\lambda_{\alpha a}^{\star}
\cdot
(|\lambda_{ba}|^{2}+|\lambda_{fa}|^{2})^{-2} \ .
\label{mea-U-Lam-4}
\end{eqnarray}
The last line in Eq.~(\ref{mea-U-Lam-4}) shows that the measure again
contains a nonintegrable singularity located at
$\mbox{ord}(|\lambda_{ba}|^{2} + |\lambda_{fa}|^{2}) = 0$.
Thus, the integral cannot be done by straightforwardly substituting 
$T_{a}=U_{a}\Lambda_{a}U_{a}^{-1}$ in $F$  and integrating with the
measure (\ref{mea-U-Lam-4}).
This can be seen by the following example.
Suppose that $F$ depends only on the
eigenvalues $\lambda_{ba},\lambda_{fa}$, and does not vanish at
$|\lambda_{ba}|^{2}+|\lambda_{fa}|^{2}$=0.
Despite the fact that the integral over $F$ in the Cartesian
coordinates is well defined, the straightforward integration in
polar coordinates gives an indefinite expression of the type
$0\cdot\infty$, with 0 and $\infty$ resulting from
the integration over the anticommuting and commuting polar variables,
respectively.
To derive the correct formula for the integration in
polar coordinates, we proceed in the following way.
We start with the Cartesian coordinates, exclude from the domain
of integration an infinitesimal neighbourhood of the singularity of 
$\mbox{d}\mu(U_{a}\Lambda_{a}U_{a}^{-1})$, and express the integral 
${\cal I}$ as the limit
\begin{equation}
{\cal I}=\lim_{\varepsilon\rightarrow 0}
\int\!\!\mbox{d}\mu(T_{a})F(T_{a})\theta(v(T_{a})-\varepsilon) \ .
\label{int-lim-4}
\end{equation}
Here $\varepsilon$ is real and positive, and $v(T_{a})$ represents the
ordinary part of
$|\lambda_{ba}|^{2} +|\lambda_{fa}|^{2}$,
as given in terms of Cartesian coordinates by
\begin{equation}
v(T_{a})
=[t_{12}^{a}]_{bb}[t_{21}^{a}]_{bb}-[t_{12}^{a}]_{ff}[t_{21}^{a}]_{ff} \ .
\label{omega-4}
\end{equation}
Since the integral on the right-hand-side of Eq. (\ref{int-lim-4}) 
is over the region where the measure
$\mbox{d}\mu(U_{a}\Lambda_{a}U_{a}^{-1})$ is regular, this integral can be
evaluated by the straightforward change to  polar coordinates.
Thus we have
\begin{equation}
{\cal I}=\lim_{\varepsilon\rightarrow 0}
\int\!\!\mbox{d}\mu(U_{a}\Lambda_{a} U_{a}^{-1})
F(U_{a}\Lambda_{a} U_{a}^{-1})
\theta(v(U_{a}\Lambda_{a} U_{a}^{-1})-\varepsilon) \ .
\label{int-pol-4}
\end{equation}
We write the function $v = v_{c} + v_{n}$ as the sum of the part
$v_{c} = |\lambda_{ba}|^{2} + |\lambda_{fa}|^{2}$ which contains
only the commuting variables $\lambda_{ba}$ and $\lambda_{fa}$, and the
nilpotent complement $v_{n}$,
\begin{eqnarray}
v_{n} = 
 (|\lambda_{ba}|^2-|\lambda_{fa}|^2)(\gamma_{1a}\gamma_{1a}^{\star}-
\gamma_{2a}\gamma_{2a}^{\star}) \nonumber \\
+2i(\lambda_{ba}\lambda_{fa}^{\star}\gamma_{2a}\gamma_{1a}^{\star}-
\lambda_{fa}\lambda_{ba}^{\star}\gamma_{1a}\gamma_{2a}^{\star}) \nonumber \\
-2 \ (|\lambda_{ba}|^2+|\lambda_{fa}|^2)\gamma_{1a}\gamma_{1a}^{\star}
\gamma_{2a}\gamma_{2a}^{\star} \ .
\label{omegc-omegn-4}
\end{eqnarray}
We expand $\theta(v_{c}+v_{n}-\varepsilon)$ in a Taylor series at
$v_{c}-\varepsilon$. The integral (\ref{int-pol-4}) decomposes into two
terms,
\begin{equation}
{\cal I}={\cal I}_{V}+{\cal I}_{W} \ .
\label{int-split-4}
\end{equation}
The volume term has the form
\begin{equation}
{\cal I}_{V}
=\lim_{\varepsilon\rightarrow 0}
\int\!\!\mbox{d}\mu(U_{a}\Lambda_{a} U_{a}^{-1})F(U_{a}\Lambda_{a} U_{a}^{-1})
\theta(v_{c}-\varepsilon) \ ,
\label{intV-4}
\end{equation}
and the boundary term or Efetov--Wegner term is given by
\begin{equation}
{\cal I}_{W}
=\lim_{\varepsilon\rightarrow 0}
\int\!\!\mbox{d}\mu(U_{a}\Lambda_{a}U_{a}^{-1})F(U_{a}\Lambda_{a}U_{a}^{-1})
(\delta(v_{c}-\varepsilon)v_{n}
+\frac{1}{2}\delta'(v_{c}-\varepsilon)v_{n}^{2}) \ .
\label{intB-4}
\end{equation}
Terms of higher order in $v_{n}$ vanish. To calculate ${\cal I}_{W}$, we
introduce new commuting coordinates $\rho,\phi,\phi_{b}$ and $\phi_{f}$ by
\begin{equation}
\lambda_{ba}=\sqrt{\rho}\cos\!\phi\,\mbox{e}^{i\phi_{b}},\;\;\;
\lambda_{fa}=\sqrt{\rho}\sin\!\phi\,\mbox{e}^{i\phi_{f}}.
\label{lamb-lamf-4}
\end{equation}
Since $v_{n}$ is linear in $\rho$, the singularity $\rho^{-1}$ in the
measure is cancelled by the factors $\rho$ and $\rho^2$ arising in the
terms due to the Taylor expansion in Eq. (\ref{intB-4}). Because of the
terms $\delta(\rho-\varepsilon)$ and $\rho\delta'(\rho-\varepsilon)$
in the integrand, the function $F(U_{a}\Lambda_{a} U_{a}^{-1})$
can be set equal to its value at $T=1$. With 
\begin{equation}
v_{n}^{2}
=-2\rho^{2}\gamma_{1a}\gamma_{1a}^{\star}\gamma_{2a}\gamma_{2a}^{\star},\quad
\int\!\!\mbox{d}\mu(U_{a})v_{n}^{j}
=-\frac{2}{(2\pi)^{2}}\rho^{j},\;\;\;j=1,2,
\label{intU-4}
\end{equation}
we find that
\begin{equation}
{\cal I}_{W}=F(1)\,\frac{1}{(2\pi)^{2}}
\lim_{\varepsilon\rightarrow 0}
\int\!\mbox{d}\rho\mbox{d}\phi\,\mbox{d}\phi_{b}\mbox{d}\phi_{f}
\sin2\phi\;\delta(\rho-\varepsilon)=F(1)\;.
\label{intB-4a}
\end{equation}
In summary, we have
\begin{equation}
{\cal I} = \lim_{\varepsilon\rightarrow 0}
\int\!\!\mbox{d}\mu(U_{a}\Lambda_{a}U_{a}^{-1})F(U_{a}\Lambda_{a}U_{a}^{-1})
\theta(|\lambda_{ba}|^2+|\lambda_{fa}|^2-\varepsilon)
+ F(1)\;.
\label{int-split-4a}
\end{equation}
This is the desired result. It expresses the integral as a volume term and
a surface contribution. 

The limit in the volume term can be taken after using the Taylor expansion
of the function $F(U_{a}\Lambda_{a}U_{a}^{-1})$ in the Grassmann variables
$\gamma_{pa}$,$\gamma_{pa}^{\star}$. Except for the term of zeroth order,
$F_{0}(U_{a}\Lambda_{a}U_{a}^{-1})$, all higher terms in the expansion tend
to zero as $\Lambda_{a}$ tends to the unit matrix. We assume that these
terms vanish as $(|\lambda_{ba}|^{2}+|\lambda_{fa}|^{2})^{\alpha}$ with
$\alpha \geq 1$. Then, only the expansion term of fourth order
$F_{4}(U_{a}\Lambda_{a}U_{a}^{-1})$ contributes. Working out the limit yields
\begin{equation}
{\cal I}={\cal I}_{V}+{\cal I}_{W}
=\int\!\!\mbox{d}\mu(U_{a}\Lambda_{a}U_{a}^{-1})
F_{4}(U_{a}\Lambda_{a}U_{a}^{-1})+F_{0}(1) \ .
\label{int-split-4b}
\end{equation}

We apply this method to an integral over a function $F(T)$ depending on 
the eight--dimensional matrix $T$,
\begin{equation}
{\cal I}=\int\!\!\mbox{d}\mu(T)F(T).
\label{int}
\end{equation}
From Subsection 2.1 it follows that the matrix $T$ is given by the product
\begin{equation}
T=U_{x\beta}\;\mbox{diag}(T_{1},T_{2})\;U_{x\beta}^{-1},
\label{T-UT12U}
\end{equation}
where
\begin{equation}
T_{r}=\left(
\begin{array}{ll}
\sqrt{1+t_{12}^{r}t_{21}^{r}}&t_{12}^{r}\\
t_{21}^{r}&\sqrt{1+t_{21}^{r}t_{12}^{r}}
\end{array}\right),\;\;\;r=1,2,
\label{Tr-mat}
\end{equation}
are the elements of the coset space
$\mbox{U}(1,1/2)/\mbox{U}(1/1)\times\mbox{U}(1/1)$
with polar coordinates $\gamma_{pr}$,$\gamma_{pr}^{\star}$ and
$\lambda_{\alpha r}$,$\lambda_{\alpha r}^{\star}$
(cf. Eqs. (\ref{U-Lam-4}) and (\ref{u12-lam-4})),
\begin{eqnarray}
T_{r}=U_{r}\Lambda_{r}U_{r}^{-1},\;\;\;
U_{r}=\left(\!
\begin{array}{ll}
u_{1r}&0\\
0&u_{2r}
\end{array}
\!\right),\;\;\;
\Lambda_{r}=\left(\!
\begin{array}{ll}
\sqrt{1+\lambda_{r}\bar{\lambda}_{r}}&\lambda_{r}\\
\bar{\lambda}_{r}&\sqrt{1+\bar{\lambda}_{r}\lambda_{r}}
\end{array}
\!\right),
\nonumber\\ \nonumber\\
u_{1r}=\exp\left(\!
\begin{array}{ll}
0&\gamma_{1r}\\
\gamma_{1r}^{\star}&0
\end{array}
\!\right),\;\;\;
u_{2r}=\exp\left(\!
\begin{array}{ll}
0&i\gamma_{2r}\\
i\gamma_{2r}^{\star}&0
\end{array}
\!\right),\;\;\;
\lambda_{r}=\left(\!
\begin{array}{ll}
\lambda_{br}&0\\
0&\lambda_{fr}
\end{array}
\!\right) \ .
\label{u12r-lamr}
\end{eqnarray}
In Eq.~(\ref{T-UT12U}), we now regard the matrices $T_{r}$ as functions
of $t_{12}^{r}$ and $t_{21}^{r}$. This defines a new parametrization of $T$
in terms of the polar coordinates of $U_{x\beta}$ and the Cartesian
coordinates of $T_{1}$ and $T_{2}$, with integration measure
\begin{equation}
\mbox{d}(U_{x\beta}\;\mbox{diag}(T_{1},T_{2})\;U_{x\beta}^{-1})
=\mbox{d}\mu(U_{x\beta})\mbox{d}\mu(T_{1})\mbox{d}\mu(T_{2})m(\Lambda).
\label{mea-UT12U}
\end{equation}
Here $\mbox{d}\mu(U_{x\beta})$ denotes the measure (\ref{mea-Uxbet}) for
integration over the matrices $U_{x\beta}$, $\mbox{d}\mu(T_{r})$ denotes
the Cartesian measures for integration over the matrices $T_{r}$
(cf. Eq. (\ref{mea-car-4})),
\begin{equation}
\mbox{d}\mu(T_{r})=\prod_{\alpha\alpha'}
\mbox{d}[t_{12}^{r}]_{\alpha\alpha'}
\mbox{d}[t_{12}^{r}]^{\star}_{\alpha\alpha'} \ ,
\label{mear-carr}
\end{equation}
and $m(\Lambda)$ denotes the factor
\begin{equation}
m(\Lambda)
=\prod_{\alpha}(|\lambda_{\alpha 1}|^{2}-|\lambda_{\alpha 2}|^{2})^{2}
\cdot \prod_{r\ne r'}(|\lambda_{br}|^{2}+|\lambda_{fr'}|^{2})^{-2}\;.
\label{muLam}
\end{equation}
This measure does not contain any nonintegrable singularity in
the domain of integration. Hence, we can write
\begin{equation}
{\cal I}=\int\!\!\mbox{d}\mu
(U_{x\beta}\;\mbox{diag}(T_{1},T_{2})\;U_{x\beta}^{-1})
F(U_{x\beta}\;\mbox{diag}(T_{1},T_{2})\;U_{x\beta}^{-1}) \ .
\label{int-UT12U}
\end{equation}
The source term (\ref{sourceterm}) vanishes at $T=1$ where $Q=L$.
Thus in the case of Eqs.~(\ref{result}) and (\ref{CnM}), the function
$F(U_{x\beta}\;\mbox{diag}(T_{1},T_{2})\;U_{x\beta}^{-1})$ 
tends to zero, $F(1)=0$, when $T_{1}$ and $T_{2}$ tend to the unit matrix.
In the integrals over $T_{1}$ and $T_{2}$, we change to polar coordinates,
using the rule (\ref{int-split-4b}). This yields
\vspace{1ex}
\begin{eqnarray}
&&{\cal I}={\cal I}_{V}+{\cal I}_{W}\;,
\label{intV-intB-f}\\ \nonumber \\
&&{\cal I}_{V}=\int\!\!
\mbox{d}\mu(U\Lambda U^{-1})F_{44}(U\Lambda U^{-1}),
\nonumber \\ \nonumber \\
&&{\cal I}_{W}=\int\!\!
\mbox{d}\mu(U_{x\beta})\mbox{d}\mu(U_{2}\Lambda_{2}U_{2}^{-1})
F_{04}(U_{x\beta}\;\mbox{diag}(1,U_{2}\Lambda_{2}U_{2}^{-1})\;
U_{x\beta}^{-1})\;.
\label{intB-f}
\end{eqnarray}
Here $F_{n_{1}n_{2}}(U\Lambda U^{-1})$ denotes the part of
$F(U\Lambda U^{-1})$ which is of order $n_{1}$ in
$\gamma_{p1},\gamma_{p1}^{\star}$ and of order $n_{2}$ in
$\gamma_{p2},\gamma_{p2}^{\star}$, and
$\mbox{d}\mu(U_{2}\Lambda_{2}U_{2}^{-1})$ denotes the integration measure
in polar coordinates for the matrices $T_{2}$, 
\begin{eqnarray}
&&\mbox{d}\mu(U_{2}\Lambda_{2}U_{2}^{-1})
=\mbox{d}\mu(U_{2})\mbox{d}\mu(\Lambda_{2}),
\nonumber \\ \nonumber \\
&&\mbox{d}\mu(U_{2})
=\prod_{p}\mbox{d}\gamma_{p2}\mbox{d}\gamma_{p2}^{\star},
\nonumber \\ \nonumber \\
&&\mbox{d}\mu(\Lambda_{2})
=\prod_{\alpha}\mbox{d}\lambda_{\alpha 2}\mbox{d}\lambda_{\alpha 2}^{\star}
\cdot (|\lambda_{b2}|^{2}+|\lambda_{f2}|^{2})^{-2} \ .
\label{mea-U-Lam-422}
\end{eqnarray}
This is the final result.

We simplify the notation by writing the integrals $\,{\cal I}_{V}$ and
$\,{\cal I}_{W}$ as
\begin{equation}
{\cal I}_{\Omega}
=\int_{\Omega}\!\mbox{d}\mu(U\Lambda U^{-1})
F(U\Lambda U^{-1})\;,\;\;\;\;\;
\Omega=V,\,W\;.
\label{int-Ome}
\end{equation}
For $\Omega=W$, the matrices $U$, $\Lambda$ and the
measure $\mbox{d}\mu(U\Lambda U^{-1})$ are by definition given by 
\begin{eqnarray}
&&U=U_{x\beta}\,\mbox{diag}(1,U_{2})\;,\;\;\;
\Lambda=\mbox{diag}(1,\Lambda_{2})\;,\;\;\;
\nonumber \\ \nonumber \\
&&\mbox{d}\mu(U\Lambda U^{-1})= 
\mbox{d}\mu(U_{x\beta})\,\mbox{d}\mu(U_{2}\Lambda_{2}U_{2}^{-1})\;.
\label{U-Lam-mea-bdr}
\end{eqnarray}
It is always understood that the only non--vanishing contribution arises
from that part of $F$ which is of the highest possible order in 
$\gamma_{pr},\gamma_{pr}^{\star}$, i.e., of 8th order for $\Omega=V$,
and of 4th order for $\Omega=W$.

\section{Asymptotic Expansion for Small $t$}
\label{asy-exp}

The behaviour of the correlator $C(t,M)=\overline{\delta g^{(1)}
\delta g^{(2)}}$ at small $t$ is described by the leading terms of the
asymptotic expansion
\begin{equation}
C(t,M)=\overline{g^{(1)}g^{(2)}}-\overline{g^{(1)}}\;\overline{g^{(2)}}
\,=\,\sum_{n=0}^{\infty}\,c(n,M)\,t^{n}-(M/2)^{2}\;,
\label{CtM}
\end{equation}
generated by expanding the exponential in the integrand of Eq.~(\ref{result})
in powers of $t$. The expansion coefficients $c(n,M)$ are given by the graded
integrals
\begin{equation}
c(n,M)={\cal N}_{n}\int\!\mbox{d}\mu(T)\,D(M,Q)\,R(Q)\,S^{n}(Q)\;,\;\;\;
Q=T^{-1}LT\;,
\label{CnM}
\end{equation}
where 
\vspace{1ex}
\begin{eqnarray}
&&{\cal N}_{n}=(-)^{n}/(2^{n}n!)\;,\;\;\;
D(M,Q)={\mbox{detg}}^{-M}(1+QL) \ ,
\nonumber \\ \nonumber \\
&&S(Q)=\mbox{trg}\,[(Q\tau_{3})^{2}]\;,
\label{DQM-SQ}
\end{eqnarray}
and where $R(Q)$ denotes the source term (\ref{sourceterm}). We note that by
definition, the channel number $M$ assumes even values only. For given $M$,
only the first $M/2+1$ terms of the expansion (\ref{CtM}) are meaningful: for
higher values of $n$, the factors $S^{n}(Q)$
cause the integrals (\ref{CnM})
to diverge (diverging integrals over the bosonic eigenvalues).
We calculate the first three terms of the expansion. The expansion coefficients
(\ref{CnM}) can be worked out analytically. The calculation relies on symbolic
manipulation utilities (we have used those of Mathematica) in an essential way
and is described in Subsection \ref{eva-coe}. The results are given in
Subsection \ref{lea-ter}.

\subsection{The Coefficients $c(n,M)$ for $n \leq 2$}
\label{eva-coe}

We change to polar coordinates, use the integration formulae
(\ref{intV-intB-f}) and (\ref{int-Ome}), and write
\begin{equation}
c(n,M)=\sum_{\Omega}c_{\Omega}(n,M)\;,
\label{CnM-COmenM}
\end{equation}
where
\begin{eqnarray}
&&c_{\Omega}(n,M)=
{\cal N}_{n}\int_{\Omega}\!\mbox{d}\mu(U\Lambda U^{-1})\,
D(M,Q)\,R(Q)\,S^{n}(Q)\;,\qquad
\nonumber\\ \nonumber\\
&&Q=U\Lambda^{-1}L\Lambda U^{-1}\;.
\label{COme-nM}
\end{eqnarray}
Both the volume part $c_{V}(n,M)$ and the boundary part $c_{W}(n,M)$ are
evaluated in the same way. We only sketch the main steps.

Following VWZ, we express the complex variables $\lambda_{\alpha r}$
in terms of absolute values and phase angles,
\begin{equation}
\lambda_{\alpha r}=i\sin\frac{1}{2}\theta_{\alpha r}\,
\mbox{e}^{i\phi_{\alpha r}}\;,\qquad
\theta_{br}=i\vartheta_{br}\;,\qquad
\theta_{fr}=\vartheta_{fr}\;,
\label{lam-the-phi}
\end{equation} 
with $\vartheta_{\alpha r}$ and $\phi_{\alpha r}$ real.
Substituting this expression in $T=U\Lambda U^{-1}$
and $Q=U\Lambda^{-1}L\Lambda U^{-1}$ yields
\vspace{1ex}
\begin{eqnarray}
&&T={\tilde U}{\tilde\Lambda}{\tilde U}^{-1}\;,\qquad
{\tilde\Lambda}=\left(
\begin{array}{ll}
\cos\frac{1}{2}\theta&i\sin\frac{1}{2}\theta\\
i\sin\frac{1}{2}\theta&\cos\frac{1}{2}\theta
\end{array}
\right)\;,
\\
&&{\tilde Q}=
{\tilde U}{\tilde\Lambda}^{-1}L{\tilde\Lambda}{\tilde U}^{-1}\;,\qquad
{\tilde\Lambda}^{-1}L{\tilde\Lambda}
=\left(
\begin{array}{ll}  
\cos\theta&i\sin\theta\\
-i\sin\theta&-\cos\theta
\end{array}
\right)\;,
\label{T-Q-til}
\end{eqnarray}
where
\begin{equation}
{\tilde U}=\mbox{diag}({\tilde u_{1}},{\tilde u_{2}})\;,\qquad
{\tilde u_{1}}=u_{1}\mbox{e}^{i\phi}\;,\qquad
{\tilde u_{2}}=u_{2}\;.
\label{U-til}
\end{equation}
The matrices ${\tilde U},{\tilde \Lambda}$ differ from their counterparts
$U,\Lambda$ only in their dependence on the phases $\phi_{\alpha r}$. We
express the matrices $T$ and $Q$ in terms of the modified matrices
${\tilde U}=\mbox{diag}({\tilde u_{1}},{\tilde u_{2}})$ and ${\tilde
\Lambda}$, and omit the tilde. It is advantageous to parametrize the
modified ``eigenvalue'' matrix $\Lambda$ in terms of 
\begin{equation}
\mu_{\alpha r}=\cos^{2}\,\frac{1}{2}\theta_{\alpha r}\;.
\label{mu-alpr}
\end{equation}
Explicit expressions for the modified matrices $U$ and $\Lambda$ and for
the associated integration measures are provided in Appendix \ref{mat-mea}.

Inverting the matrix $1+\Lambda^{-1}L\Lambda L$, we obtain
\begin{equation} 
(1+QL)^{-1}=U(1+\Lambda^{-1}L\Lambda L)^{-1}U^{-1}=
U\!\frac{1}{2}\left(\!
\begin{array}{ll}
1&i\tan\frac{1}{2}\theta\\
i\tan\frac{1}{2}\theta&1
\end{array}\!\right)\! U^{-1}\;.
\label{1+QL-inv}
\end{equation}
\vspace{1ex}
The term $D(M,Q)$ depends only on the matrix $\Lambda$,
\vspace{1ex}
\begin{equation}
D(M,Q)=
\mbox{detg}^{-M}(1+\Lambda^{-1}L\Lambda L)=D(M,\Lambda^{-1}L\Lambda)\;,
\label{DQM-DLamM}
\end{equation}
with
\begin{equation}
D(M,\Lambda^{-1}L\Lambda)\;=\;\left\{
\begin{array}{lll}
(\mu_{f1}\mu_{f2})^{M}(\mu_{b1}\mu_{b2})^{-M}
\quad&\qquad&\Omega=V\;,\\
&\mbox{for}&\\
\mu_{f2}^{M}\mu_{b2}^{-M}&&\Omega=W\;.
\end{array}
\right.
\label{DLamM-VB}
\end{equation}
Substituting Eq. (\ref{1+QL-inv}) into  the matrices
$G^{(rp)}(Q)=(1+QL)^{-1}I^{(rp)}$ of the source term $R(Q)$
and separating the parts which contain different powers of $M$,
we get
\vspace{1ex}
\begin{equation}
R(Q)=R_{2}(Q)+R_{3}(Q)+R_{4}(Q)\;,
\label{R-R234}
\end{equation}
\vspace{1ex}
where $R_{k}(Q)$, $k=2,3,4$, denote the source term components
\vspace{1ex}
\begin{eqnarray}
&&R_{2}(Q)=\frac{M^{2}}{32}\,
\mbox{trg}\Bigl[\,\chi I^{(1)}(u_{1})\chi I^{(2)}(u_{2})\,\Bigr]
\mbox{trg}\Bigl[\,\chi I^{(2)}(u_{1})\chi I^{(1)}(u_{2})\,\Bigr]\;,
\nonumber\\ \nonumber\\
&&R_{3}(Q)=\frac{M^{3}}{64}\,
\mbox{trg}\Bigl[\,\chi I^{(1)}(u_{1})\chi I^{(1)}(u_{2})
\chi I^{(2)}(u_{1}) \chi I^{(2)}(u_{2}) \nonumber \\
&&\;\;\;\;\;\;\;\;\;\;\;\;\;\;\;\;+\chi I^{(1)}(u_{1})\chi I^{(2)}(u_{2})
\chi I^{(2)}(u_{1})\chi I^{(1)}(u_{2})\,\Bigr]\;,
\nonumber\\ \nonumber\\
&&R_{4}(Q)=\frac{M^{4}}{64}\,
\mbox{trg}\Bigl[\,\chi I^{(1)}(u_{1})\chi I^{(1)}(u_{2})\,\Bigr]
\mbox{trg}\Bigl[\,\chi I^{(2)}(u_{1})\chi I^{(2)}(u_{2})\,\Bigr]\;.
\label{R234}
\end{eqnarray}
The dependence of $R_{k}(Q)$ on $U$ is contained in the graded matrices
$I^{(r)}(u_{p})$,
\begin{equation}
I^{(r)}(u_{p})=u_{p}^{-1}I^{(r)}u_{p}\;,
I^{(1)}=\mbox{diag}(1,0,-1,0)\;,
I^{(2)}=\mbox{diag}(0,1,0,-1)\;,
\label{I-I12}
\end{equation}
their dependence on $\Lambda$ in the matrix
\begin{equation}
\chi=\tan\frac{1}{2}\theta\;.
\end{equation}
Substituting Eq. (\ref{T-Q-til}) in the coupling term $S(Q)$ yields
\vspace{1ex}
\begin{equation} 
S(Q)=S_{11}(Q)+S_{22}(Q)+2S_{12}(Q)\;,
\label{S-S12}
\end{equation}
\vspace{1ex}
where $S_{pp'}(Q)$ are the graded traces
\vspace{1ex}
\begin{eqnarray}
&&S_{11}(Q) =\mbox{trg}\Bigl[\,
\cos\theta\,\tau_{3}(u_{1})\,\cos\theta\,\tau_{3}(u_{1})
\,\Bigr]\;, \nonumber \\
&&S_{22}(Q)=\mbox{trg}\Bigl[\,
\cos\theta\,\tau_{3}(u_{2})\,\cos\theta\,\tau_{3}(u_{2})
\,\Bigr]\;, \nonumber\\
&&S_{12}(Q)=\mbox{trg}\Bigl[\,
\sin\theta\,\tau_{3}(u_{1})\,\sin\theta\,\tau_{3}(u_{2})
\,\Bigr]\;,
\label{S12}
\end{eqnarray}
\vspace{1ex}
which depend on $U$ via the graded matrices 
\begin{equation}
\tau_{3}(u_{p})=u_{p}^{-1}\tau_{3}u_{p}\;,\;\;\;
\tau_{3}=\mbox{diag}(1,-1,1,-1)\;.
\label{tau3p}
\end{equation}

Writing the measure $\mbox{d}\mu(U\Lambda U^{-1})$ as the product of measures
for integration over $U$ and over $\Lambda$, using Eq.~(\ref{R-R234}) and
performing first the integration over $U$ gives the coefficients
$c_{\Omega}(n,M)$ as the sum of contributions of the three source
components,
\vspace{1ex}
\begin{equation}
c_{\Omega}(n,M)=\sum_{k}c_{\Omega}^{(k)}(n,M)\;,
\label{COmenM-COmeknM}
\end{equation}
where
\begin{equation}
c_{\Omega}^{(k)}(n,M)=
\int_{\Omega}\!\mbox{d}\mu(\Lambda)D(M,\Lambda^{-1}L\Lambda)
F_{\Omega}^{(k)}(n,M,\Lambda)\;,
\label{COmeknM}
\end{equation}
with
\begin{equation}
F_{\Omega}^{(k)}(n,M,\Lambda)=
{\cal N}_{n}\int_{\Omega}\mbox{d}\mu(U)R_{k}(Q)S^{n}(Q)\;.
\label{FOmek}
\end{equation}
The measure $\mbox{d}\mu(U)$ is the product of measures for integration
over the matrices $u_{1}$ and $u_{2}$, the source components $R_{k}(Q)$
and the traces $S_{pp'}(Q)$ depend on $u_{p}$ only in terms of the matrices
$I^{(r)}(u_{p})$ and $\tau_{3}(u_{p})$, and this dependence has the simple
form shown of Eqs.~(\ref{R234}) and (\ref{S12}). For all these reasons, the
evaluation of $F_{\Omega}^{(k)}(n,M,\Lambda)$ reduces to the calculation of
the integrals
\begin{eqnarray}
&&K_{\Omega}^{(p)}(i_{1},j_{1},i_{2},j_{2}|k_{1},l_{1}, \dots ,k_{m},l_{m})
\nonumber\\
&&=\int_{\Omega}\!\mbox{d}\mu(u_{p})
[I^{(1)}(u_{p})]_{i_{1}j_{1}}[I^{(2)}(u_{p})]_{i_{2}j_{2}}
[\tau_{3}(u_{p})]_{k_{1}l_{1}}\dots[\tau_{3}(u_{p})]_{k_{m}l_{m}}\;,
\label{XOmep}
\end{eqnarray} 
where $p = 1,2$.
From the mathematical point of view, these integrals are 
linear forms of integrals of products of matrix elements
of the coset matrices $u_{p},u_{p}^{-1}$ over the coset 
$U(2/2)/U(1)\times U(1)\times U(1)\times U(1)$,
and could therefore be evaluated by a generating function
approach similar to that worked out for the same integrals over the graded
unitary group by Guhr \cite{guh91,guh96}.

The integrals
$K_{\Omega}^{(1)}$ and $K_{\Omega}^{(2)}$  referring to the same set of
indices $i_{1},j_{1},i_{2},j_{2},$ $k_{1},l_{1},\dots ,k_{m},l_{m}$
are closely related to each other.
Starting from the integral $K_{V}^{(1)}(i_{1},j_{1},i_{2},j_{2}|k_{1},l_{1},
\dots ,k_{m},l_{m})$, making the change to the primed anticommuting
variables introduced by
$\beta_{1q}=i\beta_{1q}'$,
$\beta_{1q}^{\star}=i\tilde{\beta}_{1q}'$,
$\gamma_{1r}=i\gamma_{1r}'$,
$\gamma_{1r}^{\star}=i\tilde\gamma_{1r}'$,
and comparing the result with the integral
$K_{V}^{(2)}(i_{1},j_{1},i_{2},j_{2}|k_{1},l_{1},\dots ,k_{m},l_{m})$, 
we find that the two integrals in fact differ only by a factor, the integral
over the phases $\phi_{\alpha r}$ which is present in $K_{V}^{(1)}$ and
absent in $K_{V}^{(2)}$. Apart from an additional difference in sign due to
the Berezinian, the same is true for the integrals $K_{W}^{(1)}$ and
$K_{W}^{(2)}$. This simplifies the calculation in an essential way.
The integration over the anticommuting variables yields the part of
highest possible order of
\vspace{1ex}
\begin{equation}
(1\,-2\,\beta_{p1}\beta_{p1}^{\star}\beta_{p2}\beta_{p2}^{\star})
[I^{(1)}(u_{p})]_{i_{1}j_{1}}[I^{(2)}(u_{p})]_{i_{2}j_{2}}
[\tau_{3}(u_{p})]_{k_{1}l_{1}}\dots[\tau_{3}(u_{p})]_{k_{m}l_{m}}
\label{XOmep-intgrass}
\end{equation}
(of 16th order for $\Omega=V$, of 12th order for $\Omega=W$). After setting
$x_{p\alpha}=\sin\zeta_{p\alpha}\mbox{e}^{i\eta_{p\alpha}}$, the integration
over the commuting variables appearing in $u_{p}$ simplifies to integrals
over phase angles, or over products of powers of basic trigonometric functions. 
Multiplying the products of complementary pairs of integrals
$K_{\Omega}^{(p)}(i_{1},j_{1},i_{2},j_{2}|k_{1},l_{1},\dots ,k_{m},l_{m})$
by the $\Lambda$-dependent factors stemming from $\chi$ and $\cos\theta$ and
collecting the contributions yields the desired $F_{\Omega}^{(k)}(n,M,\Lambda)$.
The result has the form
\vspace{1ex}
\begin{eqnarray}
F_{V}^{(k)}(n,M,\Lambda)
={\cal N}_{nk}M^{k}\,\prod_{rr'}(\mu_{br}-\mu_{fr'})
\cdot\prod_{\alpha r}\mu_{\alpha r}^{-2}\cdot\,P_{V}^{(k)}(n,\Lambda)\;,
\\
F_{W}^{(k)}(n,M,\Lambda)
={\cal N}_{nk}M^{k}
(\mu_{b2}-\mu_{f2})\prod_{\alpha}\mu_{\alpha 2}^{-2}
\cdot P_{W}^{(k)}(n,\Lambda)\;,
\label{FF-PP}
\end{eqnarray}
where $P_{\Omega}^{(k)}(n,\Lambda)$ are polynomials in $\mu_{\alpha r}$,
and where
\begin{equation} 
{\cal N}_{nk}=(-)^{n}(1+\delta_{k2})/(2^{n+2}n!).
\end{equation}
The polynomials $P_{V}^{(k)}(n,\Lambda)$ are symmetrical with respect to the
interchange of $\mu_{b1}$ and $\mu_{b2}$, and of $\mu_{f1}$ and $\mu_{f2}$.
Taking into account this property, we can write the final integrals over
$\Lambda$ as
\vspace{1ex}
\begin{eqnarray}
&&c_{V}^{(k)}(n,M) \nonumber \\
&&={\cal N}_{nk}M^{k}\,\int_{V}
\mbox{d}\mu(\Lambda)\prod_{r}\mu_{fr}^{M-2}\mu_{br}^{-M-2}\cdot
\prod_{rr'}(\mu_{br}-\mu_{fr'})\cdot\,
P_{V}^{(k)}(n,\Lambda)
\nonumber \\ 
&&={\cal N}_{nk}M^{k}\,\frac{1}{4}\!
\int_{1}^{\infty}\!\!\mbox{d}\mu_{b1}\!\int_{1}^{\infty}\!\!\mbox{d}\mu_{b2}
\!\int_{1}^{0}\!\!\mbox{d}\mu_{f1}\!\int_{1}^{0}\!\!\mbox{d}\mu_{f2}\; \times
\nonumber \\
&&\quad\quad\quad\quad\frac{\prod_{r}\mu_{fr}^{M-2}}{\prod_{r}\mu_{br}^{M+2}}
\;\frac{\prod_{\alpha}\enspace(\mu_{\alpha 1}-\mu_{\alpha 2})^{2}}
{\prod_{rr'}(\mu_{br}-\mu_{fr'})}\;P_{V}^{(k)}(n,\Lambda)\;, \nonumber \\
\nonumber\\ 
&&c_{W}^{(k)}(n,M)
={\cal N}_{nk}M^{k}\,\int_{W}\mbox{d}\mu(\Lambda)\mu_{f2}^{M-2}\mu_{b2}^{-M-2}
(\mu_{b2}-\mu_{f2})\,P_{W}^{(k)}(n,\Lambda)
\nonumber \\ 
&&\quad\quad\quad = {\cal N}_{nk}M^{k}\,\int_{1}^{\infty}
\mbox{d}\mu_{b2}\!\int_{1}^{0}\mbox{d}\mu_{f2}
\frac{\mu_{f2}^{M-2}}{\mu_{b2}^{M+2}}
\frac{1}{\mu_{b2}-\mu_{f2}}P_{W}^{(k)}(n,\Lambda)\;.
\label{CV-CW}
\end{eqnarray}
The integrals can be done analytically using the approach described in
Appendix \ref{int-eig}. Collecting the contributions of all three source
components finally yields the coefficients $c_{\Omega}(n,M)$.

The calculation of the volume integrals $c_{V}^{(k)}(n,M)$ can be simplified
by using modified polar coordinates where
\begin{eqnarray}
&&T=U\Lambda U^{-1}\;,\quad U=\mbox{diag}(u_{1},u_{2})\;, \nonumber \\
&&u_{1}=u_{1\gamma}u_{1\beta}u_{1x}u_{\phi}\;,\quad
u_{2}=u_{2\gamma}u_{2\beta}u_{2x}\;,
\label{rev-par}
\end{eqnarray}
with $u_{p\gamma}$,$u_{p\beta}$,$u_{px}$ and $u_{\phi}$ having the same form
as in the old parametrization. The integration measures for both
parametrizations coincide. Since the matrices $u_{p\gamma}$ commute with
the matrix $\tau_{3}$, the matrices $\tau_{3}(u_{p})$ are now independent of
$\gamma_{pr},\gamma_{pr}^{\star}$. In the integrals $K_{V}^{(p)}$, the only
part of the integrand which then depends on these anticommuting variables is
the product of matrix elements of the matrices $I^{(r)}(u_{p})$, and only the
part of highest order in $\gamma_{pr},\gamma_{pr}^{\star}$ contributes.
Straightforward calculation shows that this part is obtained by replacing the
matrices $u_{p\gamma}^{-1}I^{(r)}u_{p\gamma}$ in $I^{(r)}(u_{p})$ by the
matrices $(-)^{p}2\gamma_{pr}\gamma_{pr}^{\star}J^{(r)}$, where the $J^{(r)}$'s
denote the projectors $J^{(1)}=\mbox{diag}(1,0,1,0)$ and 
$J^{(2)}=\mbox{diag}(0,1,0,1)$. This substitution simplifies the calculation
very much. Unfortunately, no similar coordinate transformation has been
found to simplify the boundary integrals.

\subsection{Leading Terms}
\label{lea-ter}

We begin with the expansion term of zeroth order $c(0,M)$.
The integrals $K_{\Omega}^{(p)}$ which enter the calculation do not contain
any of the matrix elements of $\tau_{3}(u_{p})$.
Integrating over
$u_{p}$ we find that $K_{V}^{(p)}(i_{1},j_{1},i_{2},j_{2}|)=0$ for all
$i_{1},j_{1},i_{2},j_{2}$. Thus $c_{V}^{(k)}(0,M)=0$ for any $k$, and the
coefficient $c(0,M)$ is entirely determined by the boundary terms
$c_{W}^{(k)}(0,M)$. We obtain, after combining the contributions from the
integrals $K_{W}^{(p)}(i_{1},j_{1},i_{2},j_{2}|)$, that
\begin{eqnarray}
&&P_{W}^{(2)}(0,\Lambda)=-\mu_{b2}+\mu_{f2}\;, \nonumber \\ 
&&P_{W}^{(3)}(0,\Lambda)=2(\mu_{b2}+\mu_{f2}-2\mu_{b2}\mu_{f2})\;, \nonumber \\
&&P_{W}^{(4)}(0,\Lambda)=P_{W}^{(2)}(0,\Lambda)\;.
\label{PB-k-0}
\end{eqnarray}
The integration over $\Lambda$ then yields
\begin{equation}
c_{W}^{(2)}(0,M)=\frac{M^{2}}{2(M^{2}-1)}=-\,c_{W}^{(3)}(0,M)\;,\;\;\;\;\;
c_{W}^{(4)}(0,M)=\frac{M^{4}}{4(M^{2}-1)}\cdot
\label{CB234-0M}
\end{equation}
The contributions $c_{W}^{(2)}(0,M)$ and $c_{W}^{(3)}(0,M)$ cancel each
other, and the coefficient $c(0,M)$ is given by the boundary contribution
of the source component $R_{4}(Q)$,
\begin{equation}
c(0,M)=c_{W}^{(4)}(0,M)-\frac{1}{4}M^{2}
=\frac{M^{2}}{4(M^{2}-1)}\cdot
\label{C0M}
\end{equation}
By definition, $c(0,M)=C(0,M)$ gives the variance $\mbox{Var}[g(B)]$ of the
dimensionless conductance $g(B)$. The value (\ref{C0M}) of $\mbox{Var}[g(B)]$
agrees with the one derived by Baranger and Mello using a random $S$--matrix
approach \cite{bar94}.

We turn to the coefficient of the term linear in $t$. Here, we meet a similar
situation: All integrals $K_{V}^{(p)}(i_{1},j_{1},i_{2},j_{2}|k_{1},l_{1})$
vanish. With
\begin{eqnarray}
&&P_{W}^{(2)}(1,\Lambda)=(8/3)\,\times \nonumber \\ 
&&\qquad\lbrace
\mu_{b2}(2-\mu_{b2}-3\mu_{b2}^{2})
-\mu_{b2}^{2}(2\mu_{b2}-3)\mu_{f2}
-\mu_{b2}^{2}(3-4\mu_{b2})\mu_{f2}^{2} \rbrace \nonumber \\
&&\qquad+(b2\leftrightarrow f2)\;, \nonumber \\
&&P_{W}^{(3)}(1,\Lambda)=-(16/3)\,\times(\mu_{b2}-\mu_{f2})
\nonumber \\
&&\qquad\lbrace
1-\mu_{b2}(1+3\mu_{b2})+2\mu_{b2}^{2}\mu_{f2}+\mu_{b2}^{2}\mu_{f2}^{2}
+(b2\leftrightarrow f2)\rbrace\;, \nonumber \\
&&P_{W}^{(4)}(1,\Lambda)=(8/3)\,\times
\nonumber \\
&&\qquad\lbrace
\mu_{b2}(2-\mu_{b2}-3\mu_{b2}^{2})
-\mu_{b2}(6-3\mu_{b2}-10\mu_{b2}^{2})\mu_{f2}
-\mu_{b2}^{2}(8\mu_{b2}-3)\mu_{f2}^{2}\rbrace \nonumber \\ 
&&\qquad+(b2\leftrightarrow f2)\;,
\label{PB-k-1}
\end{eqnarray}
the boundary contributions of the first two source components
compensate each other,
\begin{equation}
c_{W}^{(2)}(1,M)=-\frac{8M^{3}}{(M^{2}-1)^{2}}=-\,c_{W}^{(3)}(1,M),
\label{Cbdr23-1M}
\end{equation}
and the coefficient has the value
\begin{equation}
c(1,M)=c_{W}^{(4)}(1,M)=-\frac{4 M^{3}}{(M^{2}-1)^{2}}\cdot
\label{C1M}
\end{equation}

We finally turn to the term of second order in $t$. Here, both
$c_{V}(2,M)$ and $c_{W}(2,M)$ contribute because
$K_{V}^{(p)}(i_{1},j_{1},i_{2},j_{2}|k_{1},l_{1},k_{2},l_{2})$ does not vanish.
The polynomials $P_{V}^{(k)}(2,\Lambda)$ and $P_{W}^{(k)}(2,\Lambda)$
are given by
\vspace{1ex}
\begin{eqnarray}
&&P_{V}^{(2)}(2,\Lambda)=(128/9)\,\times
\nonumber\\
&&\qquad\lbrace 
\,[\,8\mu_{b1}\mu_{b2}\,\bigl(1-2(\mu_{b1}+\mu_{b2})+4\mu_{b1}\mu_{b2}
+\mu_{b1}^{2}+\mu_{b2}^{2}\,\bigr)
\nonumber\\
&&\qquad\qquad
+8(\mu_{b1}^{2}+\mu_{b2}^{2})
-16(\mu_{b1}^{3}+\mu_{b2}^{3})+12(\mu_{b1}^{4}+\mu_{b2}^{4})\,]
\,\mu_{f1}\mu_{f2}
\nonumber\\
&&\qquad -[\,\mu_{b1}\mu_{b2}\bigl(\,7(\mu_{b1}+\mu_{b2})-4(\mu_{b1}\mu_{b2}
+\mu_{b1}^{2}+\mu_{b2}^{2})\,\bigr) \nonumber \\
&&\qquad\qquad-4(\mu_{b1}^{3}+\mu_{b2}^{3})\,]\,\mu_{f1}\mu_{f2}
(\mu_{f1}+\mu_{f2})
\nonumber\\
&&\qquad -4\mu_{b1}\mu_{b2}[\,2(\mu_{b1}+\mu_{b2})-3\mu_{b1}\mu_{b2}\,]\,
(\mu_{f1}^{2}+\mu_{f2}^{2})
+36\mu_{b1}^{2}\mu_{b2}^{2}\,\mu_{f1}^{2}\mu_{f2}^{2}
\nonumber\\
&&\qquad +(b1\leftrightarrow f1,\;b2\leftrightarrow f2)\,\rbrace\;,
\nonumber \\ 
\lefteqn{
P_{W}^{(2)}(2,\Lambda)=-(64/45)(\mu_{b2}-\mu_{f2})\,\times
}\nonumber\\
\lefteqn{
\qquad\lbrace\,7-52\mu_{b2}+70\mu_{b2}^{2}-100\mu_{b2}^{3}+90\mu_{b2}^{4}
}\nonumber \\
\lefteqn{
\qquad\qquad-2\mu_{b2}(\,15-70\mu_{b2}+70\mu_{b2}^{2}-42\mu_{b2}^{3}\,)\,
\mu_{f2}
}\nonumber\\
\lefteqn{
\qquad-\mu_{b2}^{2}(\,225-404\mu_{b2}+162\mu_{b2}^{2}\,)\,\mu_{f2}^{2}
-86\mu_{b2}^{3}\,\mu_{f2}^{3}
+(b2\leftrightarrow f2)\,\rbrace\;,
} \nonumber \\
\lefteqn{
P_{V}^{(3)}(2,\Lambda)=(512/9)\,\times
}\nonumber\\
\lefteqn{
\qquad\lbrace\,[\,2\mu_{b1}\mu_{b2}(\mu_{b1}^{2}+\mu_{b2}^{2}+\mu_{b1}\mu_{b2})
+2(\mu_{b1}^{2}+\mu_{b2}^{2})
}\nonumber\\
\lefteqn{
\qquad\qquad-4(\mu_{b1}^{3}+\mu_{b2}^{3})
+3(\mu_{b1}^{4}+\mu_{b2}^{4}\,)]\,\mu_{f1}\mu_{f2}
}\nonumber\\
\lefteqn{
\qquad+[\,\mu_{b1}\mu_{b2}(\,\mu_{b1}^{2}+\mu_{b2}^{2}-8\mu_{b1}\mu_{b2}\,)
+\mu_{b1}^{3}+\mu_{b2}^{3}\,]\,\mu_{f1}\mu_{f2}(\mu_{f1}+\mu_{f2})
}\nonumber\\
\lefteqn{
\qquad+\mu_{b1}\mu_{b2}[\,2(\mu_{b1}+\mu_{b2})-3\mu_{b1}\mu_{b2}]
\,(\mu_{f1}^{2}+\mu_{f2}^{2})
}\nonumber\\
\lefteqn{
\qquad-(b1\leftrightarrow f1,\;b2\leftrightarrow f2)\,\rbrace\;,
} \nonumber \\
\lefteqn{
P_{W}^{(3)}(2,\Lambda)=(256/45)\,\times
}\nonumber\\
\lefteqn{
\qquad\lbrace\,\mu_{b2}(7-26\mu_{b2}+35\mu_{b2}^{2}
-50\mu_{b2}^{3}+45\mu_{b2}^{4})
}\nonumber\\
\lefteqn{
\qquad+\mu_{b2}(5-35\mu_{b2}+10\mu_{b2}^{2}+25\mu_{b2}^{3}-42\mu_{b2}^{4})
\mu_{f2}
}\nonumber\\
\lefteqn{
\qquad+\mu_{b2}^{2}(90-115\mu_{b2}+20\mu_{b2}^{2}-9\mu_{b2}^{3})
\,\mu_{f2}^{2}
}\nonumber\\
\lefteqn{
\qquad+5\mu_{b2}^{3}(5+3\mu_{b2})\,\mu_{f2}^{3}
+(b2\leftrightarrow f2)\,\rbrace\;,
}\nonumber \\
\lefteqn{
P_{V}^{(4)}(2,\Lambda)=P_{V}^{(2)}(2,\Lambda)\;,
}\nonumber \\
\lefteqn{
P_{W}^{(4)}(2,\Lambda)=-(2/45)(\mu_{b2}-\mu_{f2})\,\times
}\nonumber\\
\lefteqn{
\qquad\lbrace
\,7-52\mu_{b2}+70\mu_{b2}^{2}-100\mu_{b2}^{3}+90\mu_{b2}^{4}
}\nonumber \\
\lefteqn{
\qquad+2\mu_{b2}(75-230\mu_{b2}+230\mu_{b2}^{2}-138\mu_{b2}^{3})\,\mu_{f2}
}\nonumber\\
\lefteqn{
\qquad+\mu_{b2}^{2}(435-676\mu_{b2}+198\mu_{b2}^{2})\,\mu_{f2}^{2}
}\nonumber \\
\lefteqn{
\qquad+154\mu_{b2}^{3}\,\mu_{f2}^{3}
+(b2\leftrightarrow f2)\,\rbrace\;.
}
\label{P2}
\end{eqnarray}
\vspace{1ex}
The contributions of the three source components to $c_{\Omega}(2,M)$ are
\begin{eqnarray}
&&c_{V}^{(2)}(2,M)
=2{\cal N}(M^{6}-9M^{4}+18M^{2}+18)\;,
\nonumber\\
&&c_{W}^{(2)}(2,M)
=4{\cal N}(4M^{6}-12M^{4}+45M^{2}-9)\;,
\nonumber\\
\nonumber\\
&&c_{V}^{(3)}(2,M)
=2{\cal N}M^{2}(11M^{2}-39)\;,
\nonumber\\
&&c_{W}^{(3)}(2,M)
=-2{\cal N}M^{2}(9M^{4}-22M^{2}+69)\;,
\nonumber\\
\nonumber\\
&&c_{V}^{(4)}(2,M)
={\cal N}M^{2}(M^{6}-9M^{4}+18M^{2}+18)\;,
\nonumber\\
&&c_{W}^{(4)}(2,M)
=-{\cal N}M^{2}(M^{6}-18M^{4}+51M^{2}-90)\;,
\label{CV4-CB4}
\end{eqnarray}
where
\begin{equation}
{\cal N}=16/(3(M^{2}-1)^{2}(M^{2}-4)(M^{2}-9))\;.
\label{cal N}
\end{equation}
Similarly as in the terms of zeroth and first order, the total
contribution of the source components $R_{2}(Q)$ and $R_{3}(Q)$
is equal to zero,
$c^{(2)}(2,M)+c^{(3)}(2,M)=0$,
and we find that
\vspace{1ex}
\begin{equation}
c(2,M)=c_{V}^{(4)}(2,M)+c_{W}^{(4)}(2,M)
=\frac{16 M^{2}(3 M^{4}-11M^{2}+36)}{(M^{2}-1)^{2}(M^{2}-4)(M^{2}-9)}\;\cdot
\label{C2M}
\end{equation}
Collecting the terms of all three orders, we get the leading part
$C_{AE}(t,M)$ of the asymptotic expansion (\ref{CtM}) in the form
\begin{eqnarray}
C_{AE}(t,M)=\frac{M^{2}}{4(M^{2}-1)}
-\frac{4 M^{3}}{(M^{2}-1)^{2}}\;t \nonumber \\
\qquad+\frac{16 M^{2}(3 M^{4}-11M^{2}+36)}
{(M^{2}-1)^{2}(M^{2}-4)(M^{2}-9)}\;t^{2}.
\label{CtM-012}
\end{eqnarray}
(For $M=2$, only the first two terms are meaningful.)
This is our final analytic result.

\subsection{Comparison with the Lorentzian form}
\label{Com-Lor}

Efetov \cite{efe95} has shown that in the limit $M \gg 1$, the autocorrelation
function has the form of the square of a Lorentzian,
\begin{equation}
C_{SL}(t,M)=C(0,M)\left[\frac{1}{1+(t/t_{d})}\right]^{2}\;,
\label{CSL-ttd}
\end{equation}
where $t_{d}$ denotes a decay (shape) parameter. Thus, it is of interest
to compare our analytical result (\ref{CtM-012}) with the Taylor expansion
of the squared Lorentzian at $t = 0$. It is given by
\begin{equation}
C_{SL}(t,M)=\frac{M^{2}}{4(M^{2}-1)}
\left\{1-2(\frac{t}{t_{d}})+3(\frac{t}{t_{d}})^{2}+\dots\right\}\;.
\label{CSL-exp}
\end{equation}
For $M\gg 1$, the two linear terms become identical if we put $t_{d}=M/8$,
and the quadratic terms then coincide in this limit. This confirms the
expected agreement for $M\gg 1$. However, for the small channel numbers
$M$ typical for the experiments mentioned in the Introduction, no choice of
$t_d$ exists for wich the linear and quadratic terms in the expansion
of the squared Lorentzian would agree with the form (\ref{CtM-012}). Put
differently, with $t_{d}=M/8$, the decrease of the squared Lorentzian near
$t = 0$ is much smaller than that of the autocorrelation function
(\ref{CtM-012}). A numerical test for the asymptotic expansion
(\ref{CtM-012}) is presented in the next Section.

\section{Numerical Results}
\label{num-sim}

To allow for a comparison between the squared Lorentzian and the
autocorrelation function of our random matrix model also for larger values
of $t$ but small $M$, we have calculated the latter function numerically.
Here we present only the main results of this study. A more detailed
presentation of the numerical method and the results will be published
elsewhere.

After rescaling and the introduction of the now more convenient parameter
$b = k (B^{(1)} - B^{(2)}) A/ \phi_{0}\;,\quad b^{2} = 4t\;,$
the quantity of interest takes the form (cf.
Subsection~\ref{physicalassumptions})
\begin{equation}
\overline{g^{(1)}g^{(2)}}
=\int{\cal D}[H^{(I)}]{\cal D}[H^{(II)}]
\exp\left\{-\frac{N}{2}
\left[\mbox{tr}[H^{(I)}]^{2}+\mbox{tr}[H^{(II)}]^{2}\right]
\right\}g^{(1)}g^{(2)}\;.
\label{sim-ave}
\end{equation}
The conductances $g^{(r)}$ are given in terms of the $S$ matrices $S^{(r)}$.
The latter have the form
\begin{eqnarray}
&&S^{(r)}=1-2i\pi W[D^{(r)}]^{-1}W^{T}\;,\qquad
D^{(r)}=-H^{(r)}+iW^{T}W\;,
\nonumber\\
&&H^{(1)}=H^{(I)}-\frac{b}{2\sqrt N}H^{(II)}\;,\qquad\quad
H^{(2)}=H^{(I)}+\frac{b}{2\sqrt N}H^{(II)}\;, \nonumber \\
&&W_{c\mu}=\delta_{c\mu}\;,\qquad
\overline{ H_{\mu\nu}^{(C)} H_{\mu'\nu'}^{(C)} } =
\frac{1}{N} \delta_{\mu\nu'} \delta_{\mu'\nu}\;,\quad C=I,II \;.
\label{SD-H1H2W}
\end{eqnarray}
The ensemble average was performed by repeated drawings of the random
matrices $H^{(I)}, H^{(II)}$ from a random number generator. The results
for $M=2,4$ and 10, calculated with matrices of dimension $N=50$, are shown
in Fig.~1. The solid line shows the best fit by a squared Lorentzian written
in the form
\begin{equation}
C_{SL}(b,M)=C(0,M)\left[\frac{1}{1+(b/b_{d})^{2}}\right]^{2}\;,
\label{Csc}
\end{equation}
with $C(0,M)=M^{2}/[4(M^{2}-1)]$.
The values of the shape (decay) parameter $b_{d}$ are presented in Table~1.
For comparison, the table gives also the values of $b_{d}$ suggested by the
results of Efetov \cite{efe95} and Frahm  \cite{fra95}.

The table shows that the best values for $b_{d}$ are very close to Efetov's
value $\sqrt{M/2}$. For all values of $M$ considered, the autocorrelation
function $C(b,M)$ agrees rather well with the squared Lorentzian. As
expected, the agreement improves with increasing $M$. For $M=2$, the
deviations of $C(b,M)$ from the best squared--Lorentzian fit $C_{SL}(b,M)$
(with $b_{d}=1.1$) are shown in more detail in Figs.~2 and 3. Even in this
case, the deviations do not exceed $5\%$ in magnitude. For $M > 2$, the
best fit curve for $C_{SL}(b,M)$ lies within the error bars of the numerical
calculation.

As a test of the asymptotic expansion (\ref{CtM-012}), we have calculated
numerically also the second derivative $\partial^{2}C(b,M)/\partial b^{2}$
at $b=0$ for the two lowest $M$ values, $M=2$ and $M=4$.
We assumed that taking the derivatives can be interchanged with taking the
average. The derivatives of the integrand in Eq. (\ref{sim-ave}) were
analytically done by applying to the resolvents $D^{(r)}$ the formula
\begin{equation}
\frac {\partial^{n}} {\partial x^{n}} \frac {1}{A + x B}
= n! (-)^{n} \frac {1}{A + x B}
\left( B \frac {1}{A + x B} \right)^{n}\;,
\label{der-res}
\end{equation}
with $A,B$ denoting matrices independent of $x$.
Table 2 compares the second derivatives with the values derived from the
asymptotic expansion, and with the corresponding derivatives of
$C_{SL}(b,M)$.
Expressed in terms of $b$, the expansion takes the form
\begin{eqnarray}
&&C_{AE}(b,M)=\frac{M^{2}}{4(M^{2}-1)}
-\frac{M^{3}}{(M^{2}-1)^{2}}\;b^{2} \nonumber \\
&&+\frac{M^{2}(3 M^{4}-11M^{2}+36)}
{(M^{2}-1)^{2}(M^{2}-4)(M^{2}-9)}\;b^{4}.
\label{CbM-012}
\end{eqnarray}
The values derived from the asymptotic expansion agree very
well with the numerical results for both $M=2$ and $M=4$. This is not true
for the derivative of the squared Lorentzian which for $M=2$ differs
considerably from the numerical result.

The same calculation of the fourth derivative
$\partial^{4}C(0,2)/ \partial b^{4}$
yielded a divergent result. This fact caused us to analyse the second
derivative $\partial^{2}C(b,2)/$
$\partial b^{2}$ at small $|b|$ in greater detail.
The result is shown in Fig.~4. In contrast to the second derivative of the
squared Lorentzian, which at the origin rises with the second power of
$b$, the second derivative of $C(b,2)$ seems to rise with the absolute
value of $b$. The derivatives of this term linear in $|b|$ then generates
a $\delta$--function singularity in the fourth derivative. In the power
spectrum of $C(b,2)$, i.e. in the Fourier transform, such a singularity
manifests itself in an algebraic decay for large $k$.

In summary we have shown that at the $5\%$ level of accuracy, there is no
difference between our results and a squared Lorentzian even for the smallest
$M$ value, $M = 2$. Closer inspection of the autocorrelation function for
$M = 2$ and at the point $b = 0$ shows, however, that there is strong evidence
for a singularity of the fourth derivative, caused by non--analytic behavior.
This is an interesting and unexpected result for which we have no physical
explanation at present.

\section{Summary and Conclusions}
\label{conclusions}

We have investigated the magnetoconductance autocorrelation function for
ballistic electron transport through microstructures having the form of a
classically chaotic billiard. The structures were assumed to be connected
to ideal leads carrying few channels. Assuming ideal coupling between
leads and billiard, we have described this system in terms of a random
matrix model.

The autocorrelation function depends only on the field parameter $t$,
specified by the field (flux) difference, and on the channel number $M$.
Using the supersymmetry technique, we have analytically calculated the
leading terms of the asymptotic expansion of the correlation function
at small $t$. To this end, we have used integral theorems obtained by
applying Berezin's method of boundary functions. Using a generalization
of the standard polar coordinates to parametrize the coset space, we
succeeded in identifying and evaluating both volume and boundary (or
Efetov--Wegner) terms. We believe that the method developed in this paper
is of general interest for the supersymmetry technique, and we hope that
it will be helpful in  other cases.

We have shown that the first two terms of the asymptotic expansion of the
autocorrelation function are entirely given by the Efetov--Wegner terms.
This result is likely to be of general importance: Given some seemingly
very natural choice of integration variables, the boundary terms may easily
yield a or the major contribution.

For large $M$, the asymptotic expansion agrees with the corresponding small
$t$ expansion of the squared Lorentzian suggested by semiclassical theory.
For small $M$, differences exist. These were studied further by combining
our analytical work with numerical simulations. For the smallest value of
$M$, $M = 2$, the difference between the autocorrelation function and the
best squared--Lorentzian fit exists but does not exceed $5\%$ in magnitude.
This may seem irrelevant. However, a study of the
derivatives of the autocorrelation function yielded further evidence for a
statement suggested by the analytical form of the asymptotic expansion: The
autocorrelation function seems to be non--analytic in $t$ at $t = 0$. We
find this result surprising. It suggests the occurrence of non--analytic
behavior also in other correlation functions where the consequences may
even be observable. 
\vspace{1ex}

\section*{Acknowledgments}

We are grateful to Thomas Guhr for many stimulating discussions,
and to Martin Zirnbauer for advice, especially concerning the boundary
terms. Z.P. thanks the members of the Max-Planck-Institut f\"ur
Kernphysik in Heidelberg for their hospitality and support, and
acknowledges support by the Grant Agency of Czech Republic (grant
202/96/1744) and by the Grant Agency of Charles University (grants
38/97 and 142/95).

\section{Appendices}
\label{appendices}

\subsection{Matrices and Measures}
\label{mat-mea}

In the volume integrals ${\cal I}_{V}$, the matrices $U$ nad $\Lambda$
have the form
\begin{equation}
U=\left(
\begin{array}{ll}
u_{1}&0\\
0&u_{2}
\end{array}\right),\qquad
\Lambda=\left(\begin{array}{ll}
\cos\frac{1}{2}\theta&i\sin\frac{1}{2}\theta\\
i\sin\frac{1}{2}\theta&\cos\frac{1}{2}\theta
\end{array}\right),
\label{U-LamA}
\end{equation}
where $\theta$ denotes the diagonal matrix
$\theta=\mbox{diag}(i\vartheta_{b1},i\vartheta_{b2},
\vartheta_{f1},\vartheta_{f2})$,
and where $u_{1}$ and $u_{2}$ denote the matrices
\begin{equation}
u_{1}=u_{1x}u_{1\beta}u_{1\gamma}u_{\phi},\qquad 
u_{2}=u_{2x}u_{2\beta}u_{2\gamma},
\label{up}
\end{equation}
obtained by multiplying the factors (we set
$x_{p\alpha}=\sin\zeta_{p\alpha}\mbox{e}^{i\eta_{p\alpha}}$)
\vspace{1ex}
\begin{eqnarray}
&&u_{px}\!=\!\left(\!
\begin{array}{llll}
\cos\zeta_{pb}&\sin\zeta_{pb}\mbox{e}^{i\eta_{pb}}&0&0\\
-\sin\zeta_{pb}\mbox{e}^{-i\eta_{pb}}&\cos\zeta_{pb}&0&0\\
0&0&\cos\zeta_{pf}&\sin\zeta_{pf}\mbox{e}^{i\eta_{pf}}\\
0&0&-\sin\zeta_{pf}\mbox{e}^{-i\eta_{pf}}&\cos\zeta_{pf}\end{array}
\!\right)\!,\nonumber\\
&&u_{1\beta}\!=\!\left(\!
\begin{array}{llll}
1+\frac{1}{2}\beta_{11}\beta_{11}^{\star}&0&0&\beta_{11}\\
0&1+\frac{1}{2}\beta_{12}\beta_{12}^{\star}&\beta_{12}&0\\
0&\beta_{12}^{\star}&1-\frac{1}{2}\beta_{12}\beta_{12}^{\star}&0\\
\beta_{11}^{\star}&0&0&1-\frac{1}{2}\beta_{11}\beta_{11}^{\star}\end{array}
\!\right)\!,
\nonumber\\
&&u_{2\beta}\!=\!\left(\!
\begin{array}{llll}
1-\frac{1}{2}\beta_{21}\beta_{21}^{\star}&0&0&i\beta_{21}\\
0&1-\frac{1}{2}\beta_{22}\beta_{22}^{\star}&i\beta_{22}&0\\
0&i\beta_{22}^{\star}&1+\frac{1}{2}\beta_{22}\beta_{22}^{\star}&0\\
i\beta_{21}^{\star}&0&0&1+\frac{1}{2}\beta_{21}\beta_{21}^{\star}\end{array}
\!\right)\!,
\nonumber\\
&&u_{1\gamma}\!=\!\left(\!
\begin{array}{llll}
1+\frac{1}{2}\gamma_{11}\gamma_{11}^{\star}&0&\gamma_{11}&0\\
0&1+\frac{1}{2}\gamma_{12}\gamma_{12}^{\star}&0&\gamma_{12}\\
\gamma_{11}^{\star}&0&1-\frac{1}{2}\gamma_{11}\gamma_{11}^{\star}&0\\
0&\gamma_{12}^{\star}&0&1-\frac{1}{2}\gamma_{12}\gamma_{12}^{\star}\end{array}
\!\right)\!,
\nonumber\\
&&u_{2\gamma}\!=\!\left(\!
\begin{array}{llll}
1-\frac{1}{2}\gamma_{21}\gamma_{21}^{\star}&0&i\gamma_{21}&0\\
0&1-\frac{1}{2}\gamma_{22}\gamma_{22}^{\star}&0&i\gamma_{22}\\
i\gamma_{21}^{\star}&0&1+\frac{1}{2}\gamma_{21}\gamma_{21}^{\star}&0\\
0&i\gamma_{22}^{\star}&0&1+\frac{1}{2}\gamma_{22}\gamma_{22}^{\star}\end{array}
\!\right)\!,
\label{up-fac}
\end{eqnarray}
and $u_{\phi}=\mbox{e}^{i\phi}$, with
$\phi=\mbox{diag}(\phi_{b1},\phi_{b2},\phi_{f1},\phi_{f2})$.
The corresponding integration measure is
\vspace{1ex}
\begin{equation}
\mbox{d}\mu(U\Lambda U^{-1})=\mbox{d}\mu(U)\mbox{d}\mu(\Lambda)\;,
\label{mea-polA}
\end{equation}
where $\mbox{d}\mu(U)$ denotes the measure for integration over the
matrices $U$ given by
\vspace{1ex}
\begin{eqnarray}
&&\mbox{d}\mu(U)\enspace=\prod_{p}\mbox{d}\mu(u_{p})\;,
\nonumber\\
&&\mbox{d}\mu(u_{1})\enspace
=\mbox{d}\mu(u_{1x})\mbox{d}\mu(u_{1\beta})\mbox{d}\mu(u_{1\gamma})
\mbox{d}\mu(\phi), \nonumber \\
&&\mbox{d}\mu(u_{2})
=\mbox{d}\mu(u_{2x})\mbox{d}\mu(u_{2\beta})\mbox{d}\mu(u_{2\gamma})\;,
\nonumber\\ 
&&\mbox{d}\mu(u_{px})
=\prod_{\alpha}\mbox{d}\zeta_{p\alpha}\mbox{d}\eta_{p\alpha}
\sin 2\zeta_{p\alpha}\;, \nonumber \\
&&\mbox{d}\mu(u_{\phi})=\prod_{\alpha r}\mbox{d}\phi_{\alpha r}\;,
\nonumber\\
&&\mbox{d}\mu(u_{p\beta})
=\prod_{q}\mbox{d}\beta_{pq}\mbox{d}\beta_{pq}^{\star}
\cdot\prod_{p}
(1-2\beta_{p1}\beta_{p1}^{\star}\beta_{p2}\beta_{p2}^{\star})\;, \nonumber \\
&&\mbox{d}\mu(u_{p\gamma})
=\prod_{r}\mbox{d}\gamma_{pr}\mbox{d}\gamma_{pr}^{\star}\;,
\label{mea-U-u1u2}
\end{eqnarray}
and where $\mbox{d}\mu(\Lambda)$ denotes the measure for integration
over the matrices $\Lambda$,
\vspace{1ex}
\begin{equation}
\mbox{d}\mu(\Lambda)
=\prod_{\alpha r}\mbox{d}\mu_{\alpha r}
\cdot\prod_{\alpha}(\mu_{\alpha 1}-\mu_{\alpha 2})^{2}\cdot
\prod_{rr'}(\mu_{br}-\mu_{fr'})^{-2}\;,
\label{mea-LamA}
\end{equation}
with $\mu_{br}=\cosh^{2}\frac{1}{2}\vartheta_{br}$ and
$\mu_{fr}=\cos^{2}\frac{1}{2}\vartheta_{fr}$.
The domains of integration over (the ordinary parts of) $\zeta_{p\alpha}$
extend from 0 to $\pi/2$, those of integration over $\eta_{p\alpha},
\phi_{\alpha r}$ from 0 to $2\pi$, and those of integration over
$\mu_{br}$ and $\mu_{fr}$ from 1 to $\infty$ and 1 to 0, respectively,
with $\mu_{\alpha 1} < \mu_{\alpha 2}$.
However, as discussed in Subsection \ref{eva-coe}, calculating the volume
integrals can be simplified drastically by using a modified
parametrization where the matrices $u_{1}$ and $u_{2}$ are given
by the products
\begin{equation}
u_{1}=u_{1\gamma}u_{1\beta}u_{1x}u_{\phi},\qquad 
u_{2}=u_{2\gamma}u_{2\beta}u_{2x}\;.
\label{upmod}
\end{equation}
The integration measure in the modified parametrization has the same
form as in the old parametrization.

In the boundary integrals ${\cal I}_{W}$, all variables labelled by $r=1$
are to be set equal to zero: the matrices $u_{px}$ and $u_{p\beta}$ appearing
in $U$ have again the form shown in Eqs.~(\ref{up-fac}), whereas the matrices 
$u_{1\gamma},u_{2\gamma},\phi$ and $\theta$ simplify to
\vspace{1ex}
\begin{eqnarray}
&&u_{1\gamma}\!=\!\left(\!
\begin{array}{llll}
1&0&0&0\\
0&1+\frac{1}{2}\gamma_{12}\gamma_{12}^{\star}&0&\gamma_{12}\\
0&0&1&0\\
0&\gamma_{12}^{\star}&0&1-\frac{1}{2}\gamma_{12}\gamma_{12}^{\star}\end{array}
\!\right)\!, \nonumber \\
&&u_{2\gamma}\!=\!\left(\!
\begin{array}{llll}
1&0&0&0\\
0&1-\frac{1}{2}\gamma_{22}\gamma_{22}^{\star}&0&i\gamma_{22}\\
0&0&1&0\\
0&i\gamma_{22}^{\star}&0&1+\frac{1}{2}\gamma_{22}\gamma_{22}^{\star}\end{array}
\!\right)\!,
\nonumber\\ \nonumber \\
&&\phi=\mbox{diag}(0,\phi_{b2},0,\phi_{f2})\;, \nonumber \\
&&\theta=\mbox{diag}(0,i\vartheta_{b2},0,\vartheta_{f2})\;.
\label{ULamdif-W}
\end{eqnarray}
In the integration measure $\mbox{d}\mu(U\Lambda U^{-1})$, the measures for
integration over the matrices $u_{p\gamma},u_{\phi}$ and $\Lambda$ simplify to
\begin{eqnarray}
&&\mbox{d}\mu(u_{p\gamma})
=\mbox{d}\gamma_{p2}\mbox{d}\gamma_{p2}^{\star}\;, \qquad
\mbox{d}\mu(u_{\phi})
=\prod_{\alpha}\mbox{d}\phi_{\alpha 2}\;, \nonumber \\
&&\mbox{d}\mu(\Lambda)
=\prod_{\alpha}\mbox{d}\mu_{\alpha 2}
\cdot\,(\mu_{b2}-\mu_{f2})^{-2}\;.
\label{meadif-W}
\end{eqnarray}

\subsection{Integration over Eigenvalues}
\label{int-eig}

The integration over $\Lambda$ of $F_{W}^{(k)}(n,\Lambda,M)$ leads to
the two-dimensional integrals
\vspace{1ex}
\begin{equation}
{\cal I}(p2,q2)=
\int_{1}^{\infty}\mbox{d}\mu_{b2}\int_{0}^{1}\mbox{d}\mu_{f2}
\frac{\mu_{f2}^{M-2+q2}}{\mu_{b2}^{M+2-p2}}
\frac{1}{\mu_{b2}-\mu_{f2}}\;,
\label{int-p2q2}
\end{equation}
with $p2,q2$ nonnegative integers. The integrals converge if $M+2>p2$
and $M+q2>1$; since $M\geq 2$ for physical reasons, the second condition
is always satisfied. With $x=\mu_{f2}/\mu_{b2}$ and $y=\mu_{f2}$, we can
write
\begin{equation}
{\cal I}(p2,q2)=\int_{0}^{1}\mbox{d}x\;\frac{x^{M+1-p2}}{1-x}
\int_{x}^{1}\mbox{d}y\;y^{p2+q2-4}\;.
\label{int-p2q2-1}
\end{equation}
For $p2+q2\not=3$, the integration over $y$ gives
\begin{eqnarray}
{\cal I}(p2,q2)&=&\frac{1}{3-p2-q2}
\int_{0}^{1}\mbox{d}x\;\frac{x^{M-2+q2}-x^{M+1-p2}}{1-x}\nonumber\\
&=&\frac{1}{3-p2-q2}\left(\psi(M+2-p2)-\psi(M-1+q2)\right)\;,
\label{int-p2q2-2}
\end{eqnarray}
where $\psi(z)$ denotes Euler's $\psi$ function \cite{abr70}. Making use of
\begin{equation}
\psi(p+1)=\sum_{k=1}^{p}\frac{1}{k}+\psi(1)
\label{psi}
\end{equation}
yields
\begin{eqnarray}
&&{\cal I}(p2,q2)
=\frac{1}{3-p2-q2}\sum_{k=M-1+q2}^{M+1-p2}\frac{1}{k}\;\;\;\;\;\;\;
\mbox{for}\;\;\;p2+q2<3\;,
\nonumber\\ 
&&{\cal I}(p2,q2)
=\frac{1}{p2+q2-3}\sum_{k=M+2-p2}^{M-2+q2}\frac{1}{k}\;\;\;\;\;\;\;
\mbox{for}\;\;\;p2+q2>3\;.
\label{int-p2q2-4}
\end{eqnarray}
For $p2+q2=3$, the integration over $y$ leads to
\begin{equation}{\cal I}(p2,3-p2)
=-\int_{0}^{1}\mbox{d}x\frac{x^{M+1-p2}\ln x}{1-x}
=\zeta(2,M+2-p2)\;,
\label{int-p2q2-5}
\end{equation}
where $\zeta(z,q)$ denotes Riemann's $\zeta$ function \cite{gra80}.
Substituting the explicit expression, we get
\begin{equation}
{\cal I}(p2,3-p2)=\frac{\pi^{2}}{6}-\sum_{k=1}^{M+1-p2}\frac{1}{k^{2}}\;.
\label{int-p2q2-6}
\end{equation}
The integration over $\Lambda$ of $F_{V}^{(k)}(n,\Lambda,M)$ leads
to the four-dimensional integrals
\begin{eqnarray}
\nonumber\\
&&{\cal I}(p1,p2,q1,q2)
\nonumber\\
&&=\int_{1}^{\infty}\!\!\mbox{d}\mu_{b1}\int_{1}^{\infty}\!\!\mbox{d}\mu_{b2}
\int_{0}^{1}\!\!\mbox{d}\mu_{f1}\int_{0}^{1}\!\!\mbox{d}\mu_{f2}\;
\frac{\mu_{f1}^{M-2+q1}\mu_{f2}^{M-2+q2}}{\mu_{b1}^{M+2-p1}\mu_{b2}^{M+2-p2}}
\;\frac{\prod_{\alpha}\enspace(\mu_{\alpha 1}-\mu_{\alpha 2})^{2}}
{\prod_{rr'}(\mu_{br}-\mu_{fr'})}\;,
\nonumber\\
\label{int-p12q12}
\end{eqnarray}
with $p1,p2,q1,q2$ nonnegative integers. Using the identity
\begin{eqnarray}
\nonumber\\
&&\frac{\prod_{\alpha}\enspace(\mu_{\alpha 1}-\mu_{\alpha 2})^{2}}
{\prod_{rr'}(\mu_{br}-\mu_{fr'})}
=-\frac{\mu_{b1}-\mu_{b2}}{\mu_{b1}-\mu_{f1}}
-\frac{\mu_{b1}-\mu_{b2}}{\mu_{b1}-\mu_{f2}}
+\frac{\mu_{b1}-\mu_{b2}}{\mu_{b2}-\mu_{f1}}
+\frac{\mu_{b1}-\mu_{b2}}{\mu_{b2}-\mu_{f2}}
\nonumber\\
&&\qquad\qquad\qquad\quad\quad\quad
-\frac{(\mu_{b1}-\mu_{b2})^{2}}{(\mu_{b1}-\mu_{f1})(\mu_{b2}-\mu_{f2})}
-\frac{(\mu_{b1}-\mu_{b2})^{2}}{(\mu_{b1}-\mu_{f2})(\mu_{b2}-\mu_{f1})}\;,
\nonumber\\
\label{dec}
\end{eqnarray}
we can express this integral in terms of the two-dimensional integrals
(\ref{int-p2q2}) just considered, as seen from
\begin{eqnarray}
\nonumber\\
&&\int_{1}^{\infty}\!\!\mbox{d}\mu_{b1}\int_{1}^{\infty}\!\!\mbox{d}\mu_{b2}
\int_{0}^{1}\!\!\mbox{d}\mu_{f1}\int_{0}^{1}\!\!\mbox{d}\mu_{f2}\;
\frac{\mu_{f1}^{M-2+q1}\mu_{f2}^{M-2+q2}}
{\mu_{b1}^{M+2-p1}\mu_{b2}^{M+2-p2}}
\frac{\mu_{b1}-\mu_{b2}}{\mu_{b1}-\mu_{f1}}
\nonumber\\
&&=\frac{1}{(M+1-p2)(M-1+q2)}{\cal I}(p1+1,q1) \nonumber \\
&&-\frac{1}{(M-p2)(M-1+q2)}{\cal I}(p1,q1)\;,
\label{int1}
\end{eqnarray}
and
\begin{eqnarray}
&&\int_{1}^{\infty}\!\!\mbox{d}\mu_{b1}
\int_{1}^{\infty}\!\!\mbox{d}\mu_{b2}
\int_{0}^{1}\!\!\mbox{d}\mu_{f1}
\int_{0}^{1}\!\!\mbox{d}\mu_{f2}\;
\frac{\mu_{f1}^{M-2+q1}\mu_{f2}^{M-2+q2}}{\mu_{b1}^{M+2-p1}\mu_{b2}^{M+2-p2}}
\frac{(\mu_{b1}-\mu_{b2})^{2}}{(\mu_{b1}-\mu_{f1})(\mu_{b2}-\mu_{f2})}
\nonumber\\
&&\lefteqn{}\nonumber\\
&&={\cal I}(p1+2,q1)\,{\cal I}(p2,q2)-2{\cal I}(p1+1,q1)\,{\cal I}(p2+1,q2)
\nonumber \\
&&+{\cal I}(p1,q1)\,{\cal I}(p2+2,q2)\;.
\label{int2}
\end{eqnarray}

\newpage

\mbox{}



\begin{table}[h]
\caption{The value of the shape parameter $b_{d}$ for the best
squared-Lorentzian fit $C_{SL}(b,M)$ for the channel numbers
$M=2,4$ and $10$. The last two columns present
the $b_{d}$ values $\sqrt{M/2}$ and $\sqrt{(M+1)/2}$ predicted by
Efetov \cite{efe95} and Frahm \cite{fra95} respectively.}
\begin{tabbing}
$M$\hspace{3cm}\=$b_{d}$\hspace{3cm}\=$\sqrt{M/2}$\hspace{3cm}\=
$\sqrt{(M+1)/2}$ \\ \\
2 \> 1.10 \> 1.00 \> 1.22 \\
4 \> 1.44 \> 1.41 \> 1.58 \\
10 \> 2.20 \> 2.24 \> 2.35
\end{tabbing}
\end{table}

\begin{table}[h]
\caption{The value of the second derivative of the
correlation function $C(b,M)$ at $b=0$ for $M=2$ and $M=4$. 
The values obtained from the asymptotic expansion $C_{AE}(b,M)$
and from the best squared-Lorentzian fit $C_{SL}(b,M)$ are shown for
comparison.}
\begin{tabbing}
$M$\hspace{1cm}\=
$\partial^{2}C(0,M)/\partial b^{2}$\hspace{1cm}\=
$\partial^{2}C_{AE}(0,M)/\partial b^{2}$\hspace{1cm}\=
$\partial^{2}C_{SL}(0,M)/\partial b^{2}$ \\ \\
2 \> -1.76 \> -1.78 \> -1.08 \\
4 \> -0.52 \> -0.57 \> -0.50
\end{tabbing}
\end{table}




\begin{figure}[h]
\begin{center}
\mbox{\psfig{file=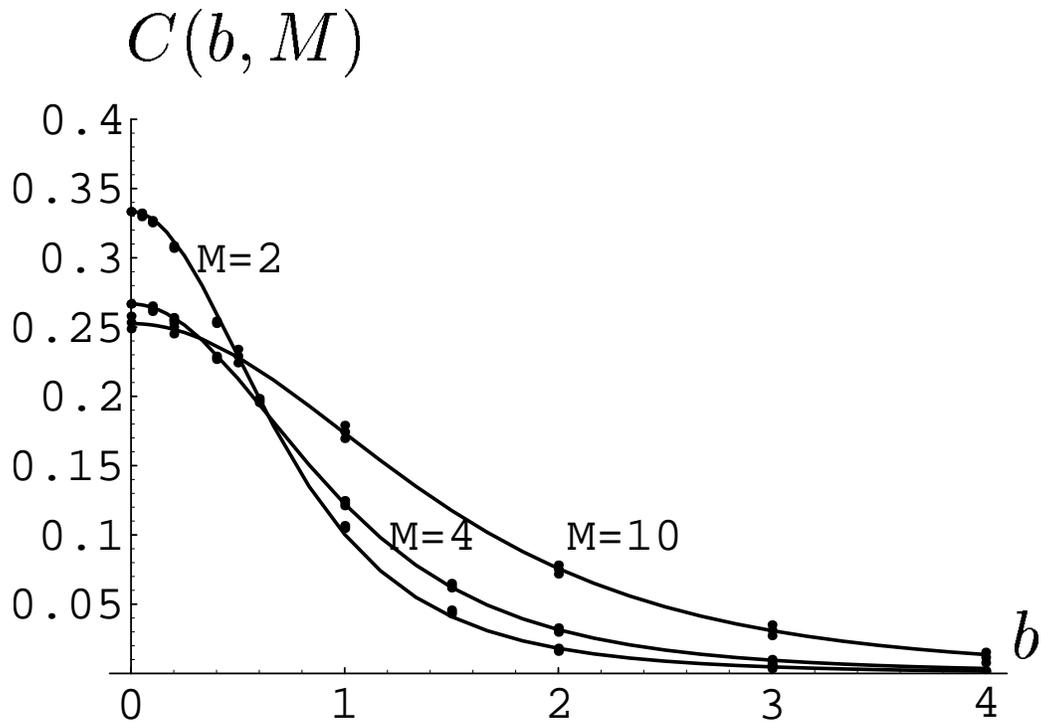,width=14cm}}
\end{center}
\caption{The correlation function $C(b,M)$ as a function of the
magnetic-field parameter $b$ for the channel numbers $M=2,4$ and $10$.
The solid lines show the corresponding  best squared--Lorentzian fit
$C_{SL}(b,M)$. 
The calculated values of the correlation function are lying in the 
middle of the error bar intervals.}
\end{figure}


\mbox{}
\begin{figure}[h]
\begin{center}
\mbox{\psfig{file=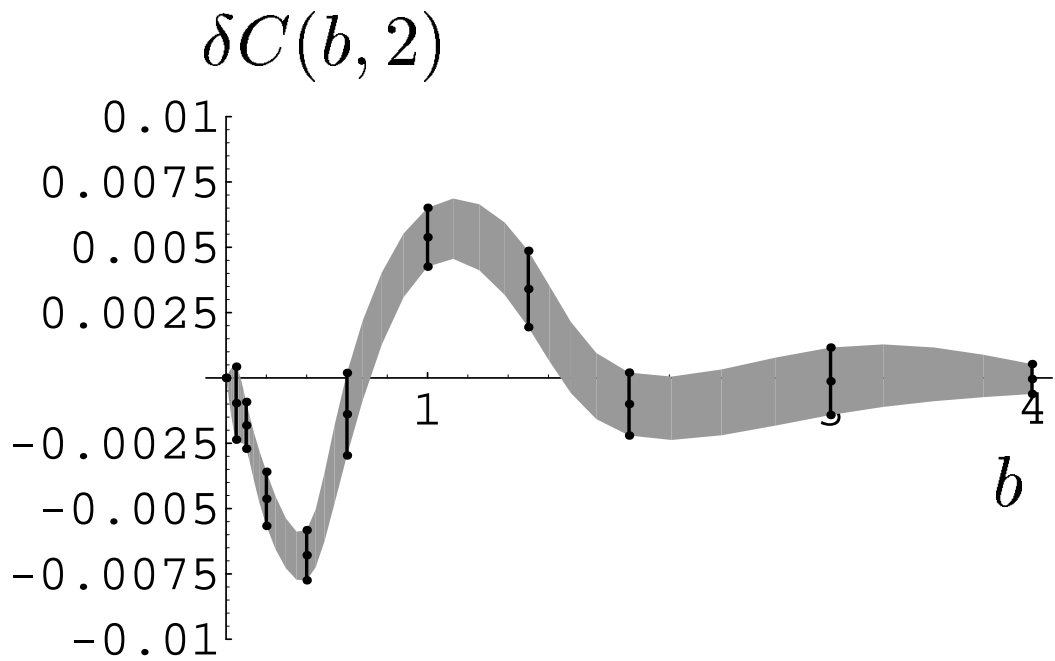,width=14cm}}
\end{center}
\caption{The deviation $\delta C(b,2) = C(b,2) - C_{SL}(b,2)$
of the correlation function $C(b,2)$ from its best squared--Lorentzian
fit $C_{SL}(b,2)$ as a function of $b$.
The gray band connecting the error bars is intended to help the eye.}
\end{figure}


\mbox{}
\begin{figure}[h]
\begin{center}
\mbox{\psfig{file=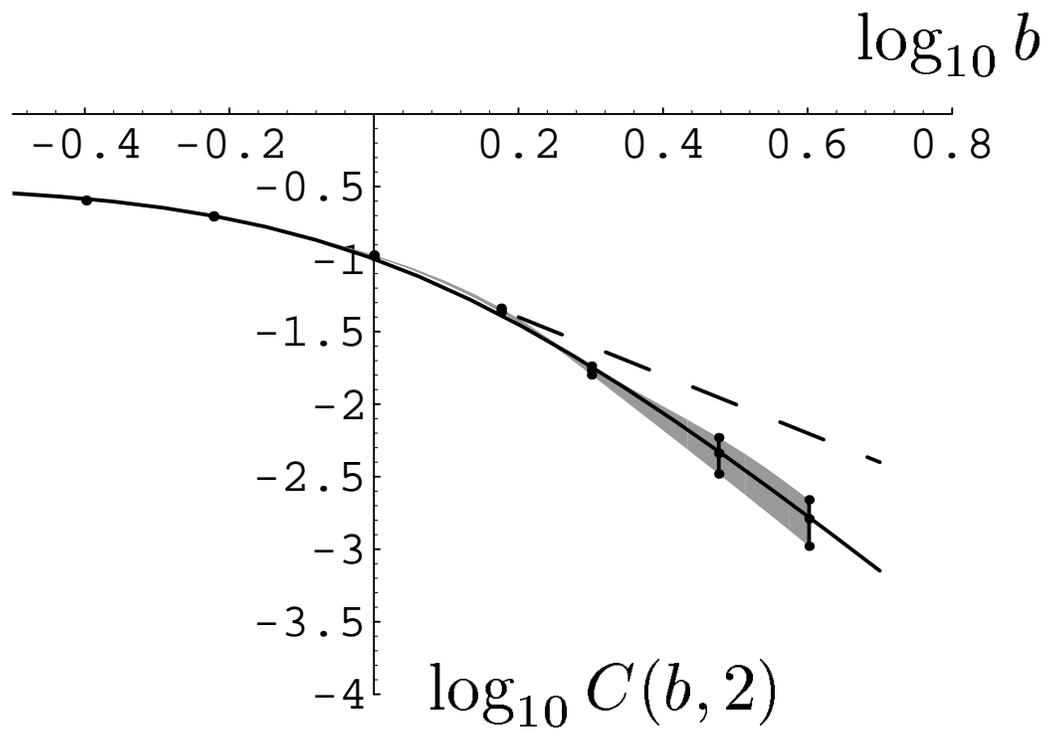,width=14cm}}
\end{center}
\caption{Plot of $\log_{10}C(b,2)$ as a function of
$\log_{10}b$.
The solid line refers to the best squared--Lorentzian fit,
the dashed line to the $b^{-2}$ law, for comparison.}
\end{figure}


\mbox{}

\begin{figure}[h]
\begin{center}
\mbox{\psfig{file=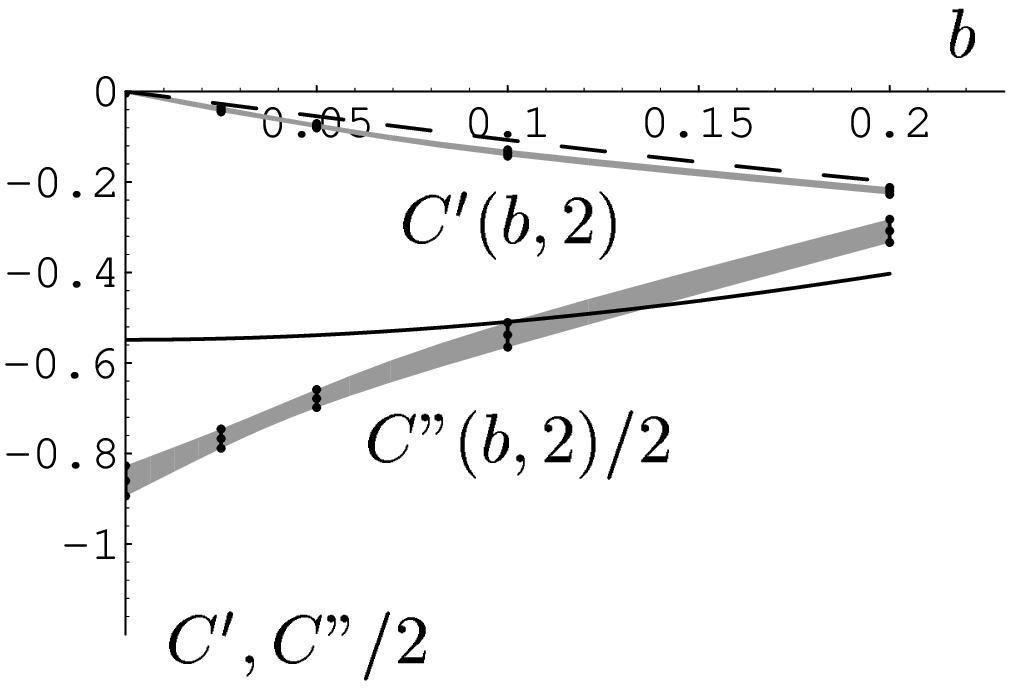,width=14cm}}
\end{center}
\caption{Graph of the first two derivatives of $C(b,2)$ as
a function of $b$.
The upper curve shows the derivative  
$C'(b,2)=\partial C(b,2)/\partial b$,
the lower curve the function 
$C''(b,2)=\partial^{2}C(b,2)/\partial b^{2}$ scaled by the factor 1/2.
The dashed and the solid line show their respective counterparts for 
the best squared--Lorentzian fit $C_{SL}(b,2)$.}
\end{figure}


\begin{thebibliography}{99}
\bibitem{ree89}
{\em Nanostructure Physics and Fabrication}, edited by M.A. Reed and
W.P. Kirk (Academic, New York 1989)
\bibitem{bee91}
C.W.J. Beenakker and H. van Houten, in {\em Solid State Physics},
edited by H. Ehrenreich and D. Turnbull (Academic, New York 1991),
Vol. 44
\bibitem{Mar92}
C.M. Marcus, A.J. Rimberg, R.M. Westerwelt, P.F. Hopkins, and A.C. Gossard,
Phys. Rev. Lett. 69(1992)506
\bibitem{cha94}
A.M. Chang, H.U. Baranger, L.N. Pfeiffer, and K.W. West,
Phys. Rev. Lett. 73(1994)2111
\bibitem{Plu95}
Z. Pluha\v r, H.A. Weidenm\"uller, J.A. Zuk, C.H. Lewenkopf, and F.J. Wegner,
Ann. Phys. 243(1995)1
\bibitem{jal90}
R.A. Jalabert, H.U. Baranger, and A.D. Stone,
Phys. Rev. Lett. 65(1990)2442
\bibitem{rau95}
J. Rau,
Phys. Rev. B51(1995)7734
\bibitem{frp95}
K. Frahm and J.-L. Pichard,
J. Phys. I France 5(1995)847
\bibitem{efe95}
K.B. Efetov,
Phys. Rev. Lett. 74(1995)2299
\bibitem{fra95}
K. Frahm,
Europhys. Lett. 30(1995)457
\bibitem{Efe83}
K.B. Efetov, Adv. Phys. 32(1983)53
\bibitem{Ver85}
J.J.M. Verbaarschot, H.A. Weidenm\"uller and M.R. Zirnbauer,
Phys. Rep. 129(1985)367
\bibitem{par79}
G. Parisi and N. Sourlas,
Phys. Rev. Lett. 43(1979)744
\bibitem{zir86}
M.R. Zirnbauer,
Nucl. Phys. B265[FS15](1986)375
\bibitem{ber87}
F.A. Berezin, {\em{Introduction to Superanalysis}} (Reidel, Dordrecht 1987)
\bibitem{rot87}
M.J. Rothstein,
Trans. Am. Math. Soc. 299(1987)387
\bibitem{zir95}
M.R. Zirnbauer and F.D.M. Haldane,
Phys. Rev. B52(1995)8729 
\bibitem{guh91}
T. Guhr,
J. Math. Phys. 32(1991)336 
\bibitem{guh96}
T. Guhr, 
Commun. Math. Phys. 176(96)555
\bibitem{bar94}
H.U. Baranger and P.A. Mello,
Phys. Rev. Lett. 73(1994)142
\bibitem{abr70}
{\em Handbook of Mathematical Functions}, edited by
M. Abramowitz and I.A. Stegun (Dover, New York 1970)
\bibitem{gra80}
I.S. Gradshteyn and I.M. Ryzhik, {\em{Table of Integrals, Series, and
Products}} (Academic, New York 1980)
\end{thebibliography}
\end{document}